\documentclass{article}

\usepackage{url}

\usepackage{natbib}
\bibpunct{(}{)}{;}{a}{}{,}

\usepackage{amssymb}

\usepackage{graphicx}
\graphicspath{{.}}

\newcommand{\sub}{\textit{Sub}}

\usepackage{fullpage}

\title{REM beyond dyads: relational hyperevent models for multi-actor interaction networks\thanks{We thank participants of the \emph{EUSN Satellite Meeting on Relational Event Models}, 6--7 September 2019, ETH Zurich for helpful comments on a preliminary version of RHEM. We acknowledge financial support from Deutsche Forschungsgemeinschaft (DFG Grant Nr.\ LE 2237/2-1),  Economic and Social Research Council (grant number ES/R009236/1), and Executive Agency for Higher Education, Research, Development and Innovation Funding (UEFISCDI grant, code PN-III-1.1-TE-2016-0362).}}

\author{
  J\"urgen Lerner\\University of Konstanz \\\url{juergen.lerner@uni-konstanz.de}
  \and
  Mark Tranmer\\ University of Glasgow \\\url{Mark.Tranmer@glasgow.ac.uk}
  \and
  John Mowbray\\ University of Glasgow \\\url{John.Mowbray@glasgow.ac.uk}
  \and
  Marian-Gabriel Hancean\\ University of Bucharest\\ \url{gabriel.hancean@sas.unibuc.ro}
}

 \date{}

\begin{document}

\maketitle

% ----------------------------------------------------------------------
\begin{abstract}
We introduce relational hyperevent models (RHEM) as a generalization of relational event models to events occurring on hyperedges involving any number of actors. RHEM can specify time-varying event rates for the full space of directed or undirected hyperedges and can be applied to model, among others, meetings, team assembly, team performance, or multi-actor communication. We illustrate the newly proposed model on two empirical hyperevent networks about meetings of government ministers and co-authoring of scientific papers.

\end{abstract}
% ----------------------------------------------------------------------

%\tableofcontents

% ----------------------------------------------------------------------
\section{Introduction}
\label{sec:intro}

Relational event models (REM) \citep{b-refsa-08,bls-ness-09,lbsb-mftien-13,sb-iat-17} are a general framework for modeling networks of time stamped interaction events. REM specify time-varying event rates associated with dyads comprising a source (the sender of the event) and a target (the receiver of the event). The source and target nodes are often ``elementary'' units representing, for instance, individuals or objects but can also represent predefined subsets of elementary nodes \citep{b-refsa-08}, such as a whole school class which may receive broadcast messages from the teacher \citep{dubois2013hierarchical}. Yet, in many situations interaction events can occur on arbitrary subsets of nodes, that is, on \emph{hyperedges} rather than on edges with exactly two endpoints \citep{kim2018hyperedge}. For instance, meetings may involve any number of persons, scientific papers may be written by any number of authors, emails may be sent to any number of recipients. Treating such hyperedges as collections of independent (dyadic) edges is invalid in general \citep{chodrow2019configuration}.

The predominant approach to model the rate (or probability) of directed or undirected \emph{hyperevents} seems to be to specify dyadic event rates and let hyperedges assemble as a function of these dyadic event rates \citep{guimera2005team,kim2018hyperedge}. This approach could overestimate or underestimate the true event rates associated with hyperedges involving more than two nodes. For instance, Alice may frequently send emails to her friend Bob and may also send many emails to her boss Charlie; yet the probability that Alice sends the same email to both, Bob and Charlie, may be close to zero. Such higher-order dependencies cannot be expressed by models restricted to specifying dyadic event rates.

To overcome limitations of previous models, this paper makes the following methodological contributions:
\begin{enumerate}
\item We propose \emph{relational hyperevent models (RHEM)} as a general framework for networks of undirected and directed relational hyperevents. RHEM, which are defined as a straightforward generalization of REM, can specify event rates on the full space of all possible hyperedges.
\item We apply well-established approximation methods to overcome the prohibitive runtime for evaluating the full likelihood function, which grows exponentially in the number of nodes. By doing so we succeed in reliably estimating RHEM parameters from a co-authoring network comprising hundreds of thousands of nodes and hyperevents. 
\item We propose and discuss a range of \emph{hyperedge statistics} that characterize how arbitrary subsets of nodes are embedded into the network of past events and that shape the rate of future hyperevents. These hyperedge statistics express hypothetical effects in hyperevent networks which can be tested with empirical data. 
\item We discuss differences between models that condition on the observed size of hyperevents and models that control for the size of hyperevents. Moreover, we propose a restriction of RHEM to repeated events and discuss why this restriction seems necessary and appropriate when modeling extremely sparse hyperevent networks.
\item We illustratively apply RHEM to two empirical hyperevent networks -- where events represent meetings and co-authored papers, respectively -- and make suggestions for best practices in modeling relational hyperevent networks of varying size.
\item We propose \emph{relational outcome models} (ROM) which explain outcome of relational hyperevents (rather than the occurrence of events) and apply them to model team performance in a co-authoring network.
\end{enumerate}

RHEM are introduced in Sect.~\ref{sec:rhem}, applied to empirical data on meetings in Sect.~\ref{sec:meetings} and to co-author networks in Sect.~\ref{sec:coauthor}. We discuss methodological insights in Sect.~\ref{sec:discussion} and further related work in Sect.~\ref{sec:related}. We end with concluding remarks and an outline of future work.

% ----------------------------------------------------------------------
\section{Relational hyperevent models (RHEM)}
\label{sec:rhem}

% ----------------------------------------------------------------------
\subsection{Background and notation: hypergraphs and relational hyperevents}
\label{sec:notation}

As a generalization of graphs, hypergraphs contain hyperedges which can connect any number of nodes; see, e.\,g., \citet{bretto2013hypergraph} for a general treatment of hypergraphs. We recall that an \emph{undirected hypergraph} is a pair $G=(V,H)$, where $V$ is a finite set of nodes and $H\subseteq 2^V=\mathcal{P}(V)=\{V'\subseteq V\}$ is a set of \emph{undirected hyperedges}. An undirected hyperedge $h\subseteq V$ is a set of any number of nodes from $V$ and the \emph{size} or \emph{cardinality} of an undirected hyperedge $h\in H$, denoted by $|h|$, is its number of elements. An undirected, loopless graph can be seen as an undirected hypergraph in which every hyperedge has size two.

A \emph{directed hypergraph} is a triple $G=(U,V,H)$, where $U$ and $V$ are finite sets of nodes (which may be identical) and $H\subseteq 2^U\times 2^V$ is a set of \emph{directed hyperedges}. Each directed hyperedge is a pair $h=(a,b)\in H$ where $a\subseteq U$ is a set of any number of \emph{source nodes} and $b\subseteq V$ is a set of any number of \emph{target nodes}. A directed hyperedge is a \emph{loop} if its set of sources has a non-empty intersection with its set of targets. The \emph{size} or \emph{cardinality} of a directed hyperedge $h=(a,b)$ is  the pair of integers $|h|=(|a|,|b|)$ giving the number of sources and the number of targets, respectively. A directed graph can be seen as a directed hypergraph in which every hyperedge has size $(1,1)$.

Given a set of nodes $V$, an \emph{undirected hyperevent} is a tuple
\[
e=(h_e,t_e,x_e,y_e)\enspace,
\]
where $h_e\subseteq V$ is an undirected hyperedge giving the \emph{participants} of the event, $t_e$ is the time of the event, $x_e$ is the \emph{type} and/or \emph{weight} of the event, and $y_e$ is a \emph{relational outcome} of the event. Types, weights, and/or relational outcome may be absent in some settings. The difference between $x_e$ and $y_e$ lies purely in the time order: the type $x_e$ may have an influence on the occurrence of the event, whereas the relational outcome $y_e$ is a result of the event. To illustrate this, assume that we are given hyperevents representing teams of actors jointly performing a given task.
In this setting, the type of an event could be the type of the task which can determine that the event is more likely to occur on one set of participants than on another (actors may have a preference to perform certain types of tasks). In the same setting, an example of a relational outcome would be given by the success or performance of the team work. The success cannot have an influence on team assembly (that is, on the hyperedge of the event) since the success is unknown at the time when team members are selected. Conversely, the team selection (that is, the hyperedge of the event) can have an influence on the success as some teams may have a higher performance than others. We further note that the relational outcome of an event at time $t$ may have an influence on the selection of the hyperedge of a future event occurring at time $t'>t$ since, for instance, actors may be inclined to collaborate with others that have a history of prior success.

The definition of directed hyperevents is very similar: given two sets of nodes $U$ and $V$ (which may be identical), a \emph{directed hyperevent} is a tuple
\[
e=(h_e,t_e,x_e,y_e)\enspace,
\]
where $h_e=(a_e,b_e)$ is a directed hyperedge with $a_e\subseteq U$ and $b_e\subseteq V$ giving the \emph{sources} and \emph{targets} of the event, $t_e$ is the time of the event, $x_e$ is the \emph{type} and/or \emph{weight} of the event, and $y_e$ is a \emph{relational outcome} of the event.

The \emph{size} or \emph{cardinality} of a hyperevent is the size of the underlying hyperedge. The size of hyperevents may be constrained in some application settings. For instance, email communication events typically have exactly one sender and an unconstrained number of receivers.

A set of nodes $h'$ is called a \emph{sub-hyperedge} of an undirected hyperedge $h$ if $h'$ is a subset of $h$, that is, if $h'\subseteq h$. A pair of sets of nodes $h'=(a',b')$ is called a \emph{sub-hyperedge} of a directed hyperedge $h=(a,b)$ if $a'\subseteq a$ and $b'\subseteq b$. We denote the set of all sub-hyperedges of a (directed or undirected) hyperedge $h$ by $\sub(h)$,\footnote{For an undirected hyperedge $h$, we might also use the established notation for the power set $\mathcal{P}(h)$ or $2^h$, i.\,e., the set of all subsets of $h$. We prefer to denote sub-hyperedges by $\sub(h)$ since it provides a unified notation for undirected and directed hyperedges.} the set of all sub-hyperedges of size $k$ of an undirected hyperedge $h$ by $\sub^{(k)}(h)$, and the set of all sub-hyperedges of size $(k,l)$ of a directed hyperedge $h$ by $\sub^{(k,l)}(h)$. Sub-hyperedges are important for specifying RHEM because if a hyperevent occurs on some hyperedge $h$, then the actors in every sub-hyperedge $h'\subset h$ experience a common event (although not exclusively), which in turn may influence the probability that $h'$ is included in the hyperedge of a future event -- potentially with other co-participants.

% ----------------------------------------------------------------------
\subsection{General model specification}
\label{sec:model}

Let $E=(e_1,\dots,e_N)$ be a sequence of (undirected or directed) hyperevents, given in non-decreasing order in time. We propose two types of model frameworks for specifying the probability (density) of $E$: the first framework, which is a straightforward generalization of REM \citep{b-refsa-08}, explains the occurrence of events and the second explains the relational outcome. We defer the description of relational outcome models to Sect.~\ref{sec:rom} and introduce in this section models explaining the occurrence of hyperevents. The definition of the model framework is almost identical for directed and undirected hyperevents; however, differences become more pronounced when we define concrete statistics for hyperevents.

For a time point $t$ let $R_t$ denote the \emph{risk set} that is the set of hyperedges on which a hyperevent could happen at time $t$. For instance, the risk set may contain all subsets of a given set of nodes. Let $G[E;t]$ denote the \emph{network of past events} \citep{bls-ness-09} at time $t$ which is a function of $E_{<t}=\{e\in E\,;\;t_e<t\}$ that is of all events that happen strictly before $t$ (and potentially exogenously given covariates).

For a hyperedge $h\in R_t$ and a point in time $t$, let $T$ be the random variable for the time of the next event on $h$. The \emph{event rate} (also denoted as \emph{hazard rate} or \emph{intensity}) on $h$ at time $t$, given the network of past events is defined by
% ----------------------------------------------------------------------
\begin{equation}\label{eq:lambda_def}
  \lambda(h;t;G[E;t]) =
\lim_{\Delta t\to 0}\frac{Prob(t\leq T \leq t+\Delta t\,|\;t\leq T;G[E;t])}{\Delta t}
\enspace.
\end{equation}
% ----------------------------------------------------------------------

The probability (density) of the sequence of hyperevents $E$ can be speficied in various ways. In this paper we specify the likelihood function based on the Cox proportional hazard model \citep{cox1972regression} -- corresponding to the ``ordinal'' model of \citet{b-refsa-08} -- which explains relative event rates. Several alternatives for modeling time-to-event data exist \citep{l-smmld-03}.

Following the Cox proportional hazard model, we decompose the event rate $\lambda$ into a time-varying baseline rate $\lambda_0(t)$, which is the same for all hyperedges, and a relative event rate $\lambda_1(h;t;\theta;G[E;t])$ which is proportional to the probability that the event at time $t$ occurs on the hyperedge $h$, rather than on any other hyperedge in the risk set:
% ----------------------------------------------------------------------
\begin{eqnarray}
  \label{eq:lambda}
  \lambda(h;t;\theta;G[E;t])&=&\lambda_0(t)\cdot \lambda_1(h;t;\theta;G[E;t])
  \enspace,\\
  \label{eq:lambda_one}
  \lambda_1(h;t;\theta;G[E;t])&=&
  \exp\left(\theta\cdot s(h;t;G[E;t])\right)\enspace.
\end{eqnarray}
% ----------------------------------------------------------------------
In the equations above, $s(h;t;G[E;t])$ is a vector of \emph{statistics} characterizing how the hyperedge $h$ is embedded into the network of past events and $\theta$ is a vector of parameters. 
The baseline rate $\lambda_0$ is typically left unspecified, or estimated by non-parametric methods, and the partial likelihood based on the observed event sequence $E$ is
% ----------------------------------------------------------------------
\begin{equation}
  \label{eq:likelihood}
  L(\theta)=\prod_{e\in E}\frac{\lambda_1(h_{e};t_{e};\theta;G[E;t_{e}])}
  {\sum_{h\in R_{t_{e}}}\lambda_1(h;t_{e};\theta;G[E;t_{e}])}\enspace.
\end{equation}
% ----------------------------------------------------------------------

We emphasize how close this framework for relational hyperevents is to REM \citep{b-refsa-08,perry2013point,vpr-remslm-15}. Virtually the only difference is that separate event rates are specified for all hyperedges in the risk set, that is, potentially for all subsets of a given set of nodes. This difference, however, has strong implications for the computational complexity to evaluate the likelihood function and for the range of possibilities to define statistics for hyperedges.

% ----------------------------------------------------------------------
\subsection{Model estimation and case-control sampling}
\label{sec:estimation}

Given the values of the statistics $s(h;t_{e};G[E;t_{e}])$ for all elements of the risk sets $R_{t_{e}}$ at the event times $t_{e}$, maximum likelihood estimates for Eq.~(\ref{eq:likelihood}) can be computed with standard statistical software, such as the R \texttt{survival} package\footnote{\url{https://CRAN.R-project.org/package=survival}} \citep{therneau2013modeling}. The problem in doing so is the prohibitive size of the risk set which scales exponentially with the number of nodes. In our first empirical case study presented in Sect.~\ref{sec:meetings} we analyze data on some 800 meeting events potentially involving any subset of a fixed set of 23 persons. The risk set size is $2^{23}\approx 8$ million which might still be manageable -- although the computational effort seems excessive for such a small network. In the second case study presented in Sect.~\ref{sec:coauthor}, we analyze data on more than 300\,000 papers (corresponding to hyperevents) written by authors from a pool of more than 500\,000 researchers. The maximum number of authors per paper is 100, which was chosen as a limit. Even with the limit on the size of hyperedges, we get a prohibitive risk set size of more than ${500\,000}\choose{100}$ subsets.

The problem arising from the size of the risk set is much more severe for relational hyperevent models than for REM. In the latter, the risk set grows quadratically in the number of nodes which might already result in an infeasible runtime for larger networks. \citet{b-refsa-08} proposed to mitigate such problems by sampling from the risk set, \citet{vpr-remslm-15} applied case-control sampling \citep{bgl-mascdcphm-95} in which for each event (``case'') a fixed number of ``controls'' (non-events from the risk set) are sampled, \citet{lerner2019reliability} performed an experimental reliability study of case-control sampling in the estimation of dyadic REM, found that models can be fitted to data with a risk set size of more than 30 trillion, and proposed to experimentally assess the parameter variability by repeated sampling.

Encouraged by these results, we propose to also apply case-control sampling when estimating relational hyperevent models. Note that the risk set size of our first empirical study is much smaller than the one considered in \citet{lerner2019reliability} but the size in the second study is larger by a huge factor.

For a fixed number $m$ of controls per event, we consider for each event $e\in E$ a sampled risk set $\tilde{R}_{t_{e}}$ which contains the hyperedge $h_e$ plus $m$ additional hyperedges uniformly and independently drawn at random from $R_{t_{e}}$ without replacement. The sampled likelihood $\tilde{L}$ is obtained by replacing the full risk sets in Eq.~(\ref{eq:likelihood}) by the sampled risk sets:
% ----------------------------------------------------------------------
\begin{equation}
  \label{eq:sampled_likelihood}
  \tilde{L}(\theta)=\prod_{e\in E}\frac{\lambda_1(h_{e};t_{e};\theta;G[E;t_{e}])}
  {\sum_{h\in \tilde{R}_{t_{e}}}\lambda_1(h;t_{e};\theta;G[E;t_{e}])}\enspace.
\end{equation}
% ----------------------------------------------------------------------
\citet{lerner2019reliability} propose to additionally sample from the observed events, which can be a strategy to scale models to even more events.

% ----------------------------------------------------------------------
\subsection{Network effects}
\label{sec:effects}

Network effects are specified via hyperedge statistics which are real-valued functions characterizing how hyperedges at a given time $t$ are embedded into the network of past events $G[E;t]$. These statistics are used in the specification of the relative event rate in Eq.~(\ref{eq:lambda_one}) and therefore determine the distribution of future events. Possibilities to define hyperedge statistics are abundant. In the following we describe in a modular way (1) how past events and/or exogenous covariates are aggregated into hyperedge attributes and (2) how hyperevent statistics are defined as a function of hyperedge attributes. Many variants exist for each step which often can be freely combined.

% ----------------------------------------------------------------------
\subsubsection{Hyperedge attributes}
\label{sec:attributes}

For an undirected hypergraph with an underlying node set $V$, a \emph{hyperedge attribute} is a partial function $\textit{att}\colon\sub(V)\to\mathbb{R}$ defined on hyperedges among the nodes from $V$. For a directed hypergraph with an underlying set $U$ of possible senders and a set $V$ of possible receivers, a \emph{hyperedge attribute} is a partial function $\textit{att}\colon\sub((U,V))\to\mathbb{R}$ defined on hyperedges connecting some or all senders from $U$ to some or all receivers from $V$. Hyperedge attributes are \emph{partial} functions since they may be defined only on a subset of these hyperedges, for instance, only on hyperedges of a given size. 

Hyperedge attributes encode the state of the network of past events $G[E;t]$ at a given time $t$. They can be functions of exogenously given covariates or can be functions of past events. For the latter kind we mention two types of hyperedge attributes that are particularly important and whose definition is identical for undirected and directed hyperevents.

\paragraph{Hyperedge activity} gives for a hyperedge $h$ and a point in time $t$ the number of hyperevents on $h$ that happen strictly before $t$.
\[
\textit{hyperedge.activity}(h;t;E)=\sum_{e\in E_{<t}}\chi(h = h_e)\enspace.
\]
(The function $\chi(\cdot)$ is the indicator function that is one if the argument is true and zero else.) 

\paragraph{Hyperedge degree} gives for a hyperedge $h$ and a point in time $t$ the number of hyperevents on any superset of $h$ that happen strictly before $t$.
\begin{equation}\label{eq:degree}
\textit{hyperedge.degree}(h;t;E)=\sum_{e\in E_{<t}}\chi(h\subseteq h_e)\enspace.
\end{equation}
For instance, for a one-element hyperedge $h=\{v\}$, with $v\in V$, the hyperedge degree at time $t$ is the number of past hyperevents in which $v$ participated. For a two-element hyperedge $h=\{u,v\}$, with $u,v\in V$, the hyperedge degree at time $t$ is the number of past hyperevents in which $u$ and $v$ co-participated. In general, for a set of nodes $h\subseteq V$, the hyperedge degree at time $t$ is the number of past hyperevents in which all nodes from $h$ co-participated. Statistics based on the hyperedge degree operationalize effects in which events are \emph{partially repeated} in the sense that some, but not necessarily all, co-participants of past events jointly experience future events, potentially with other co-participants.

The difference between hyperedge activity and hyperedge degree is that hyperedge activity requires past hyperevents to happen exactly on the focal hyperedge $h$, without any further participants outside of $h$. In contrast, the definition of hyperedge degree counts past hyperevents in which all participants of the focal hyperedge $h$ co-participated, but allows that these hyperevents involved further participants outside $h$.

\paragraph{Variations.} Rather than summing up the number of past hyperevents, hyperedge attributes may also take into account (functions of) the type, weight, or relational outcome of past events, may be dichotomized, or may decay the contribution of past events over time, similar as this has been suggested for dyadic events \citep{lbsb-mftien-13}. A further variant for the hyperedge degree involves down-weighting co-participation in \emph{large} events. For example, for two actors $u$ and $v$, the experience to co-participate in an event $h$ might be stronger if $h$ is small (since then it is a rather ``exclusive'' event) than if $h$ is large (in which case $u$ and $v$ might not even be aware that they co-participate in the same event). Down-weighting the impact of large events on the hyperedge degree can, for instance, be done by dividing the terms in Eq.~(\ref{eq:degree}) by the size of the hyperedge $h_e$.

% ----------------------------------------------------------------------
\subsubsection{Hyperedge statistics}
\label{sec:statistics}

The relative event rate in Eq.~(\ref{eq:lambda_one}) has been specified as a parametric function of hyperedge statistics $s(h;t;G[E;t])$, characterizing how the hyperedge $h$ is embedded into the network of past events $G[E;t]$. In the following, we define a list of hyperedge statistics, most of which are inspired by related statistics which are typically used in REM \citep{b-refsa-08,bls-ness-09,lbsb-mftien-13,sb-iat-17}.

\paragraph{Hyperedge size.} For undirected hyperedges, the statistic \emph{size} gives the number of participants.
\[
\textit{size}(h;t;G[E;t])=|h|\enspace.
\]
For a directed hyperedge $h=(a,b)$, the size is characterized by two values, the number of sources and the number of targets.
\begin{eqnarray*}
  \textit{num.sources}(h;t;G[E;t])&=&|a|\\
  \textit{num.targets}(h;t;G[E;t])&=&|b|\enspace.
\end{eqnarray*}
If the risk sets $R_t$ contain hyperedges of varying size (for instance, all hyperedges among a given set of nodes), the size of hyperedges often has a very strong impact on the event rate and therefore it is important to control for the effect of size. In our first empirical application, presented in Sect.~\ref{sec:meetings}, we find a curvilinear, U-shaped effect of size on the event rate: small hyperevents but also large hyperevents are over-represented while events occuring on hyperedges of intermediate size are relatively rare. Such a U-shaped effect can be specified by including the size and the squared size in the model statistics. Hyperedge size is one of the few hyperedge statistics that have no related statistic in dyadic REM.

\paragraph{Repetition} models the tendency to repeat hyperevents involving the identical set of participants (or identical sets of sources and targets, respectively) and is formally defined by
\[
\textit{repetition}(h;t;G[E;t])=\textit{hyperedge.activity}(h;t;E)\enspace.
\]
The definition is identical for undirected and directed hyperevents.

\paragraph{Sub-repetition (undirected).} In the case of undirected hyperevents, sub-repetition of order $p$ models the tendency of sets of $p$ nodes to repeatedly experience joint events, potentially within larger sets of participants. It is formally defined by 
\[
\textit{sub-repetition}^{(p)}(h;t;G[E;t])=\frac{1}{{|h| \choose p}}
\sum_{h'\in\sub^{(p)}(h)}\textit{hyperedge.degree}(h';t;E)\enspace.
\]
The statistic as defined above takes the average hyperedge degree over all sub-hyperedges $h'\subseteq h$ of size $p$. For $p=1$ it gives the average number of previous hyperevents involving the individual participants of $h$, for $p=2$ it gives the average number of previous joint events of all dyads among the participants of $h$, for $p=3$ it considers previous joint events of all triads contained in $h$, and so on.

Rather than the average we might also use other functions to aggregate values on all sub-hyperedges, such as the minimum, maximum, sum, or standard deviation. For instance, taking the standard deviation would give the dispersion of the hyperedge degree over the sub-hyperedges of $h$.

The definition above counts a past hyperevent $e$ occurring on a hyperedge $h_e$ as often as there are $p$-element subsets of the intersection $h_e\cap h$, that is, precisely $|h_e\cap h| \choose p$ times. This may put too much weight on hyperevents having large intersections with the focal hyperedge $h$. An alternative definition of sub-repetition -- denoted by \emph{shared prior events of order $p$} -- is to count past hyperevents only once if their intersection with $h$ has at least $p$ elements; in formulas:
\[
\textit{shared.prior.events}^{(p)}(h;t;G[E;t])=\sum_{e\in E_{<t}}\chi(|h_e\cap h|\geq p)\enspace.
\]
Variations of this alternative definition can take type, weight, or relational outcome of past events into account, may decay the impact of past events over time, or may normalize the statistic by the size of the hyperedge $h$.

\paragraph{Sub-repetition (directed).} In the case of directed hyperevents, sub-repetition of order $(p,q)$ models the tendency of a set of $p$ sources (potentially together with other source nodes) to repeatedly send events to a set of $q$ targets (and potentially other target nodes). For a hyperedge $h=(a,b)$, it is formally defined by 
\[
\textit{sub-repetition}^{(p,q)}(h;t;G[E;t])=\frac{1}{{|a| \choose p}\cdot {|b| \choose q}}
\sum_{h'\in\sub^{(p,q)}(h)}\textit{hyperedge.degree}(h';t;E)\enspace.
\]
This statistic takes the average hyperedge degree over all sub-hyperedges $h'\subseteq h$ of size $(p,q)$. That is, it takes the average over the combinations of all subsets $a'\subseteq a$ of $p$ sources with all subsets $b'\subseteq b$ of $q$ targets.
As in the undirected case, we might also use other functions to aggregate values on all sub-hyperedges, such as the minimum, maximum, sum, or standard deviation, or we might count shared prior events of order $(p,q)$ only once.

\paragraph{General sub-hyperedge characteristics.} We note that the approach to aggregate values over all sub-hyperedges (or all sub-hyperedges of a given size) is not restricted to hyperedge degrees but can be applied to any hyperedge attribute, irrespective of whether this attribute represents exogenous covariates or is a function of past events. For instance, assuming that we are given node-level attributes -- that is, attributes defined on hyperedges of size one -- representing, for instance, age or past performance of actors, we can characterize a hyperedge by the average (or any other function of) attribute values of its participants.

\paragraph{Reciprocation} models, in the case of directed hyperevents, the tendency to reciprocate events, that is, the tendency that the former set of targets becomes the set of sources, sending events directed to the former set of sources. For a directed hyperedge $h=(a,b)$ reciprocation is formally defined by
\[
\textit{reciprocation}(h;t;G[E;t])=\textit{hyperedge.activity}((b,a);t;E)\enspace.
\]
Reciprocation is only defined in one-mode hyperevent networks, where the set of potential sources is identical with the set of potential targets. The same comment applies also to the other variants of reciprocation defined below.

\paragraph{Sub-reciprocation.} In the case of directed hyperevents, sub-reciprocation of order $(p,q)$ models the tendency of sets $a'$ of $p$ former sources to receive events from sets $b'$ of $q$ former targets to which the nodes from $a'$ (potentially together with other sources) have previously sent events (potentially within larger sets of targets). For a directed hyperedge $h=(a,b)$, it is formally defined by 
\[
\textit{sub-reciprocation}^{(p,q)}(h;t;G[E;t])=\frac{1}{{|b| \choose p}\cdot {|a| \choose q}}
\sum_{h'\in\sub^{(p,q)}((b,a))}\textit{hyperedge.degree}(h';t;E)\enspace.
\]
This statistic takes the average hyperedge degree over all sub-hyperedges $h'\subseteq (b,a)$ of size $(p,q)$ of the reverse hyperedge $(b,a)$. 

We note that the parameters $p$ or $q$ in directed sub-repetition or sub-reciprocation can also be zero. Concretely, $\textit{sub-repetition}^{(p,0)}$ models a generalized repetition effect in which the same set of $p$ nodes repeatedly initiates events towards any set of targets and $\textit{sub-repetition}^{(0,q)}$ models a generalized repetition effect in which the same set of $q$ nodes repeatedly receives events from any set of sources.
Similarly, $\textit{sub-reciprocation}^{(p,0)}$ models a generalized reciprocation effect in which a set of $p$ nodes that have jointly sent events to any set of targets receives common events from any set of sources and $\textit{sub-reciprocation}^{(0,q)}$ models a generalized reciprocation effect in which a set of $q$ nodes that have received common events from any set of sources jointly sends events to any set of targets.

\paragraph{Switch reciprocation.} In the case of directed hyperevents, switch reciprocation is related to -- but more permitting than -- reciprocation. Switch reciprocation models patterns in which a part of former targets become sources and send events to former sources and former targets. Switch reciprocation is characterized by a number $l$ giving the number of nodes switching the role of source and target. Switch reciprocation is typical, for instance, in email communication where the ``reply-to-all'' functionality enables one former target (i.\,e., an actor among the receivers of a past email) to send an email to the former source (sender of that past email) and to all the other targets. This reply-to-all pattern is captured by switch reciprocation of order~$1$. For a positive integer $l$, we first define the operator $\textit{switch}^{(l)}$ mapping a directed hyperedge $h=(a,b)$ to the set of all hyperedges obtained from $h$ by interchanging $l$ elements from $a$ with $l$ elements from $b$. For simplicity we define switch reciprocation only for loopless hypergraphs. Formally the operator $\textit{switch}^{(l)}$ is defined by 
\[
\textit{switch}^{(l)}(h)=\left\{h'=(a',b')\,;\;
\exists a''\in{a \choose l},\,b''\in{b \choose l}\colon
a'=a\cup b''\setminus a'' \wedge b'=b\cup a''\setminus b''\right\}
\]
Switch reciprocation of order $l$ is formally defined by 
\begin{eqnarray*}
\textit{switch-reciprocation}^{(l)}(h;t;G[E;t])&=&\frac{1}{{|a| \choose l}\cdot {|b| \choose l}}\cdot
\sum_{h'\in\textit{switch}^{(l)}(h)}\textit{hyperedge.degree}(h';t;E)\enspace.
\end{eqnarray*}

\paragraph{Closure (undirected).} Triadic closure refers to patterns in which events on a hyperedge $h$ depend on previous events in which subsets of the nodes in $h$ have co-participated with common third actors. Triadic closure in hyperevent networks is characterized by a triple of positive integers $(p,q,l)$ where $p$ and $q$ give the sizes of two disjoint subsets $h_1,h_2\subset h$ and $l$ gives the size of a set of nodes $V'\subset V\setminus (h_1\cup h_2)$ which may (but does not have to) be disjoint from $h$ and with which nodes in $h_1$ and in $h_2$ co-participated in common events. For a triple of positive integers $(p,q,l)$ and an undirected hyperedge $h$ from a given set of nodes $V$, we define the operator $\sub^{(p,q,l)}(h)$ by
\begin{eqnarray*}
  \sub^{(p,q,l)}(h)=\{(h_1,h_2,V')&;&|h_1|=p,\,|h_2|=q,\,|V'|=l,\,h_1,h_2\subset h,\,V'\subset V,\\
  &&h_1\cap h_2=h_1\cap V'=h_2\cap V'=\emptyset\}
\end{eqnarray*}
Closure of order $(p,q,l)$ is formally defined by 
\begin{eqnarray*}
  \textit{closure}^{(p,q,l)}(h;t;G[E;t])&=&
  \frac{1}{{|h| \choose p}\cdot {|h|-p \choose q}\cdot{|V|-(p+q) \choose l}}\times\\
    &&\sum_{(h_1,h_2,V')\in\sub^{(p,q,l)}(h)}
    \min[\textit{h.deg}(h_1\cup V';t;E),\textit{h.deg}(h_2\cup V';t;E)]\enspace,
\end{eqnarray*}
where we write \textit{h.deg} as an abbreviation for \textit{hyperedge.degree}.

\paragraph{Closure (directed).} Closure in directed hyperevent networks is similar but has more variants. As in the case of dyadic relational events, we can distinguish the four variants: transitive closure, cyclic closure, shared senders, and shared receivers. Due to the simple observation that there are no triangles in two-mode networks, we define triadic closure in directed hyperevent networks only for one-mode networks where the set of potential senders $U$ is equal to the set of potential receivers $V$ and, for simplicity, we restrict the definition to loopless hypergraphs. For a triple of positive integers $(p,q,l)$ and a directed hyperedge $h=(a,b)$ from a given set of nodes $V$, we define the operator $\sub^{(p,q,l)}(h)$ by
\begin{eqnarray*}
  \sub^{(p,q,l)}(h)=\{(a',b',V')&;&|a'|=p,\,|b'|=q,\,|V'|=l,\,a'\subseteq a,\,b'\subseteq b,\,V'\subset V,\\
  &&a'\cap V'=b'\cap V'=\emptyset\}
\end{eqnarray*}
The four variants of triadic closure are formally defined by:
\begin{eqnarray*}
  \textit{transitive.closure}^{(p,q,l)}(h;t;G[E;t])&=&
  \frac{1}{{|a| \choose p}\cdot {|b| \choose q}\cdot{|V|-(p+q) \choose l}}\times\\
    &&\sum_{(a',b',V')\in\sub^{(p,q,l)}(h)}
    \min[\textit{h.deg}((a', V');t;E),\textit{h.deg}((V',b');t;E)]\\
  \textit{cyclic.closure}^{(p,q,l)}(h;t;G[E;t])&=&
  \frac{1}{{|a| \choose p}\cdot {|b| \choose q}\cdot{|V|-(p+q) \choose l}}\times\\
    &&\sum_{(a',b',V')\in\sub^{(p,q,l)}(h)}
    \min[\textit{h.deg}((V',a');t;E),\textit{h.deg}((b',V');t;E)]\\
  \textit{shared.receivers}^{(p,q,l)}(h;t;G[E;t])&=&
  \frac{1}{{|a| \choose p}\cdot {|b| \choose q}\cdot{|V|-(p+q) \choose l}}\times\\
    &&\sum_{(a',b',V')\in\sub^{(p,q,l)}(h)}
    \min[\textit{h.deg}((a', V');t;E),\textit{h.deg}((b',V');t;E)]\\
  \textit{shared.senders}^{(p,q,l)}(h;t;G[E;t])&=&
  \frac{1}{{|a| \choose p}\cdot {|b| \choose q}\cdot{|V|-(p+q) \choose l}}\times\\
    &&\sum_{(a',b',V')\in\sub^{(p,q,l)}(h)}
    \min[\textit{h.deg}((V',a');t;E),\textit{h.deg}((V',b');t;E)]
\end{eqnarray*}

\subsection{Dependencies implied by statistics}
\label{sec:implied_dependencies}

Different hyperedge statistics imply dependencies among subsets of different order. To illustrate this for the statistics sub-repetition and repetition, assume that, in the context of a co-author network, there are three different groups, each containing three actors, denoted by $\{A_i,B_i,C_i\}$ for $i=1,2,3$. Moreover, assume that each of the three actors in Group~1 has published two single-author papers, that the actors in Group~2 have published three pairwise co-authored papers (that is, three papers authored by $\{A_2,B_2\}$, $\{A_2,C_2\}$, and $\{B_2,C_2\}$, respectively), and that the third group has jointly published one three-author paper and, in addition, each of its actors has published one single-author paper.

Given these histories of past interaction, the future event rate to publish a joint three-author paper for the three hyperedges $\{A_1,B_1,C_1\}$, $\{A_2,B_2,C_2\}$, $\{A_3,B_3,C_3\}$ depends on which hyperedge statistics are included in the model. A first model depending only on sub-repetition of order one would assign the same rate to all three hyperedges. After all, each of the nine actors has published two previous papers and, thus, has the same degree (that is, the same number of previous events). This model would miss that the authors of Group~2 and Group~3 have collaborated before, but those of Group~1 did not. A second model depending on sub-repetition of order one and sub-repetition of order two would recognize the difference between Group~1 and Group~2 (the actors of Group~1 have never collaborated before, but those of Group~2 did) but would fail to distinguish between Group~2 and Group~3 in which each single author has authored the same number of papers (namely two) and each pair of authors has co-authored the same number of papers (namely one). A third model depending on sub-repetition of order one, two, and three could also distinguish between Group~2 and Group~3, since the hyperedge $\{A_3,B_3,C_3\}$ has experienced a common event before but the hyperedge $\{A_2,B_2,C_2\}$ did not. Finally, a fourth model depending only on the repetition statistic could assign different event rates to the third group $\{A_3,B_3,C_3\}$ than to Groups 1 and 2 (since Group~3 previously experienced a common event, but Groups~1 and~2 did not) but could not distinguish between $\{A_1,B_1,C_1\}$ and $\{A_2,B_2,C_2\}$ (since neither of these two hyperedges experienced any previous event.

The discussion above also illustrates that the statistic repetition and the statistics sub-repetition of order $p\geq 3$ introduce dependencies that cannot be expressed in models specifying only dyadic event rates.

% ----------------------------------------------------------------------
\subsection{Conditioning on hyperedge size}
\label{sec:conditional_size}

As it will become apparent in the empirical applications, sequences of observed hyperevents often have a very different distribution of hyperedge sizes than hyperedges randomly sampled from the set of all subsets. As a consequence of this, the hyperedge size often has a very strong impact on event rates and model fit. Seen from a different angle, defining the risk sets $R_t$ to be the entire set of all hyperedges among a given set of nodes returns an abundance of hyperedges that have an unrealistic size.

For these reasons, it might be a valuable alternative to restrict the risk sets associated with observed events $e$ to those hyperedges having the same size as $e$. In other words, we also consider models that condition on the size of the observed hyperedges. It turns out in our empirical analysis that models conditioning on the event size can produce different findings than unconstrained models. We will discuss in Sect.~\ref{sec:meetings} how these different results can be interpreted with respect to each other.
    
% ----------------------------------------------------------------------
\subsection{New events vs.\ repeated events}
\label{sec:repeated}

The probability that a hyperedge $h$ experiences any event is typically by many orders of magnitude smaller than the conditional probability that $h$ experiences a repeated event, given that there has been at least one event on $h$ before. Especially in larger networks, the marginal probability of new events is negligible compared to the conditional probability of repeated events. This observation, which is rooted in the enormous size of the risk set, can lead to numerical instability when estimating the effect of the repetition statistic -- but it also raises the conceptual question whether ``first events'' (that is, events happening on a hyperedge that has never experienced any event before) are influenced by the same network effects than repeated events. For these reasons, we consider the alternative to analyze first events on any hyperedge by a different model than repeated events. We note that repeated events, that is, events occurring on a hyperedge that has experienced at least one event before, is not the only possibility to identify a subset from the risk set that has a much higher event rate than the average rate over the unconstrained risk set. Another possibility would be, for instance, to consider only those hyperedges in which all subsets of a given order (such as, all pairs of nodes) experienced at least one event before, potentially within a larger set of participants. We discuss the rationale for modeling new events and repeated events separately in more detail in Sects.~\ref{sec:coauthor} and~\ref{sec:discussion}.

% ----------------------------------------------------------------------
\subsection{Relational outcome models}
\label{sec:rom}

Relational outcome models (ROM) explain the outcome $y_e$ of a relational hyperevent $e=(h_e,t_e,x_e,y_e)$, given that the event $e$ occurs. In settings in which hyperevents represent tasks solved by teams, a relational outcome can measure, for instance, the performance of the team on this given task. In many settings, the basic model for the outcome variable falls into the family of generalized linear models (GLM). For instance, a numeric success indicator might be modeled by linear regression or a binary indicator distinguishing between success and failure might be modeled by logistic regression. With this in mind, ROM seem to add little over the basic statistical toolbox. However, the contribution of ROM is that -- in contrast to GLM -- they can cope with certain forms of non-independence among observations. For instance, the success probability of a given hyperedge $h$ (that is, of a given team of actors) at a hyperevent at time $t$ is likely to depend on prior success of the same team $h$ -- but is also likely to depend on prior success of the individual actors in $h$, on prior shared success of dyads or triads of actors in $h$, and so on. ROM can model relational outcome dependent on the history of previous events, in the same way as RHEM model event rates dependent on the history of previous events, 

We assume that we have chosen a distribution $f(y)$ for the variable $y$ giving relational outcomes of hyperevents. ROM specify the likelihood of an observed sequence of relational events $E$, by assuming conditional independence of relational outcomes, given the network of previous events:
% ----------------------------------------------------------------------
\begin{equation}
  \label{eq:likelihood_rom}
  L(\theta)=\prod_{e\in E}f(y_{e}\,|\;h_e;t_{e};x_e;\theta;G[E;t_{e}])\enspace.
\end{equation}
% ----------------------------------------------------------------------
The distribution $f(y_{e}\,|\;h_e;t_{e};x_e;\theta;G[E;t_{e}])$ is specified as a function of parameters $\theta$ and statistics that are functions of the network of past events $G[E;t_{e}]$, as the ones defined in Sect.~\ref{sec:statistics}.

% ----------------------------------------------------------------------
\section{Case study I: Meetings}
\label{sec:meetings}

The first empirical case study that we consider in this paper is about a sequence of meeting events involving any number of participants from a relatively small set of actors. 

% ----------------------------------------------------------------------
\subsection{Setting and data}
\label{sec:meetings_setting}

% ----------------------------------------------------------------------
\begin{figure}
  \begin{center}
    \includegraphics[width=0.7\textwidth]{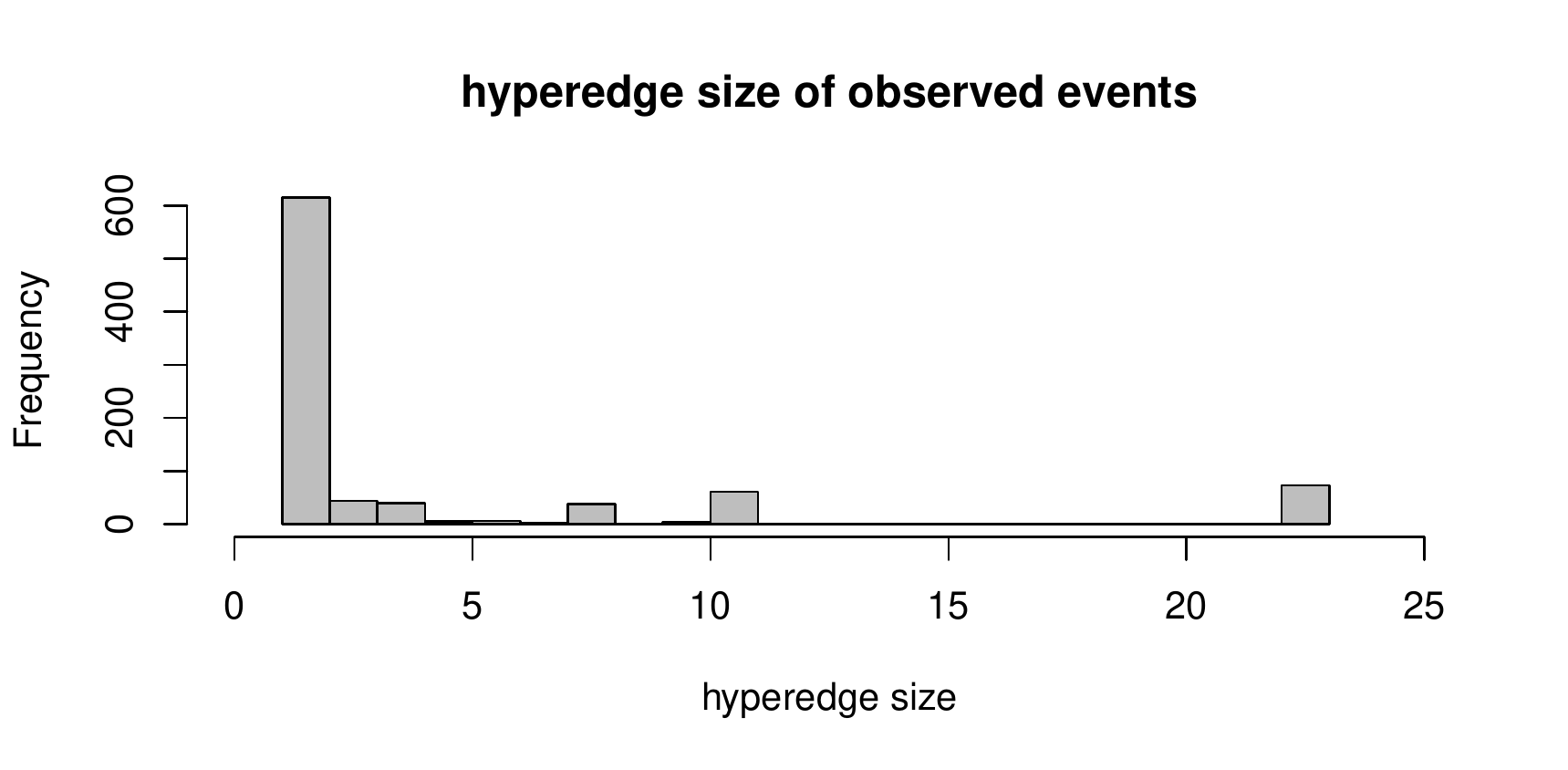}
    \includegraphics[width=0.7\textwidth]{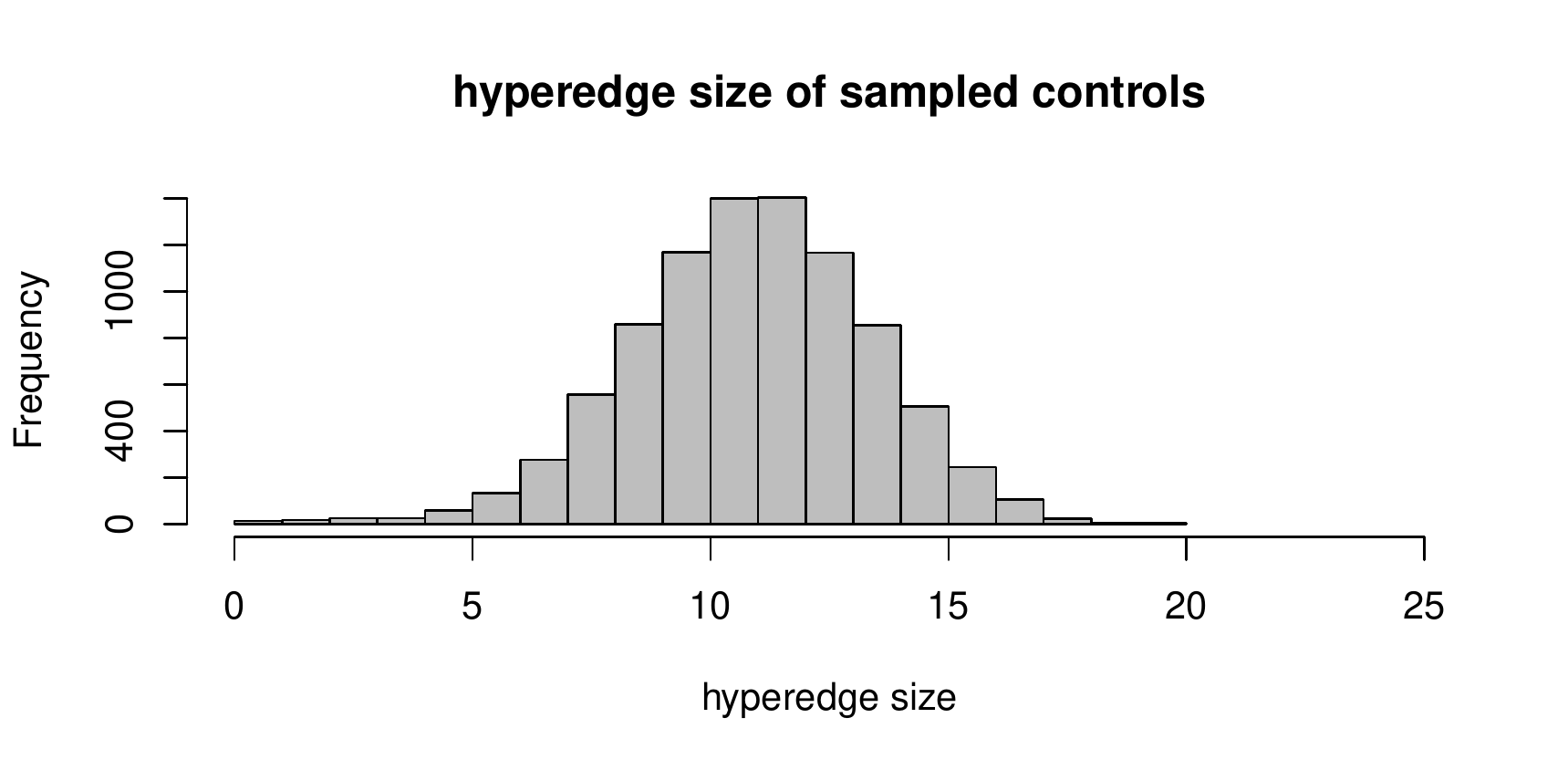}
  \end{center}
  \caption{PM meetings data. Histogram of sizes of observed hyperevents (\emph{top}) and histogram of sizes of sampled non-event hyperedges (\emph{bottom}).}
  \label{fig:pm_sizes}
\end{figure}
% ----------------------------------------------------------------------

The data for this case study are sourced from the engagement diaries of former UK Prime Minister (PM) Margaret Thatcher \citep{mtf2019}. Diary entries list the dates, times, and participants of scheduled meetings undertaken by the PM on a day-to-day basis. Here, we focus on the hyperevents (i.\,e., meetings) involving one, several, or all cabinet ministers ($n=23$) between 5th May 1979 and 31st December 1980 -- i.\,e., the beginning of Thatcher's first term as PM.\footnote{The official cabinet contained 21 ministers. However, Michael Jopling (Chief Whip) and Norman Fowler (Minister for Transport) participated in weekly cabinet meetings during the time period in focus, and their presence is represented in the data.} Although other actors often participated in these meetings, they were removed from the data in order to focus explicitly on the PM's interactions with cabinet colleagues.

In total we analyze 886 meeting events, which are listed by the minute and do not overlap. As the PM was present at each of the meetings -- and therefore her participation rate does not vary -- the underlying hyperedges comprise only of cabinet ministers, excluding the PM. Accordingly, event sizes range from 1 to 23. The size distribution of observed hyperevents -- and that of the non-event hyperedges sampled from the unconstrained risk set -- is given in Fig.~\ref{fig:pm_sizes}. This shows that small events and large events are over-represented, while events of intermediate size are relatively rare -- especially when compared to the high frequency of intermediate hyperedge sizes in the risk set. Note that the binomial distribution $f(k)={n\choose k}$ assumes a maximum at the integer $k$ closest to $n/2$. There are no types, weights, or relational outcome variables associated with these meeting events.

% ----------------------------------------------------------------------
\subsection{Model specification}
\label{sec:meetings_models}

When analyzing data from the first case study we want to tackle three issues which are mostly of methodological interest to increase our understanding of RHEM: (1) to find out whether there is any higher-order dependence in emprical hyperevent data (compare the discussion in Sect.~\ref{sec:implied_dependencies}), (2) to assess the impact of the hyperedge size (or functions of it) on the event rate, and (3) to compare models considering unconstrained risk sets with models conditioning on the observed event sizes.

To tackle these questions we first estimate a family of twelve models obtained by systematically adding the following types of hyperedge statistics in the Cox-proportional hazard models given in Eq.~(\ref{eq:sampled_likelihood}). We consider models with and without the repetition statistic and with and without the two statistics giving the size of the hyperedge and the squared size of the hyperedge. (Note that Fig.~\ref{fig:pm_sizes} suggests a curvilinear, U-shaped effect of event size on event rates, which might be reproduced by considering a polynomial of order two of the hyperedge size.) For each of the four resulting combinations, we consider a family of three models successively including sub-repetition of order one, two, and three, respectively. Results of these models are reported in Table~\ref{table:PM_uncond_size}. Besides the estimated parameters, their standard errors, and significance levels, we also report the AIC as a measure of model fit (recall that lower AIC points to better model fit).

We estimate a second family of six different models that condition on the size of observed hyperevents, as it has been discussed in Sect.~\ref{sec:conditional_size}. That is, the risk set associated with an observed hyperevent $e=(h_e,t_e)$ is defined to be $\sub^{(|h_e|)}(V_{t_e})$, that is the set of all subsets of size $|h_e|$ of the set of actors $V_{t_e}$, where the latter set comprises all cabinet ministers participating in any event at or before $t_e$. When we condition on the size of the observed hyperevent, we cannot use hyperedge size (or any function of it) as an explanatory variable since it is identical for the event and all associated non-event hyperedges. We therefore consider six models, with and without the repetition statistic and successively including sub-repetition of order one, two, and three, respectively. Results of these models are reported in Table~\ref{table:PM_cond_size}.

% ----------------------------------------------------------------------
\subsection{Results}
\label{sec:meetings_results}

\begin{table}
\begin{center}
\begin{tabular}{l c c c }
\hline
repetition.order.1 & $3.77 \; (0.15)^{***}$ & $10.60 \; (0.49)^{***}$ & $10.02 \; (0.51)^{***}$ \\
repetition.order.2 &                        & $-5.08 \; (0.27)^{***}$ & $-2.63 \; (0.40)^{***}$ \\
repetition.order.3 &                        &                         & $-2.18 \; (0.30)^{***}$ \\
\hline
AIC                & 2789.68                & 1723.40                 & 1646.03                 \\
\hline
\hline
repetition.order.1 & $4.13 \; (0.43)^{***}$  & $6.21 \; (0.84)^{***}$  & $8.77 \; (1.65)^{***}$  \\
repetition.order.2 &                         & $-1.93 \; (0.61)^{**}$  & $3.88 \; (1.90)^{*}$    \\
repetition.order.3 &                         &                         & $-8.18 \; (1.38)^{***}$ \\
repetition         & $23.12 \; (1.74)^{***}$ & $19.47 \; (1.97)^{***}$ & $13.98 \; (1.79)^{***}$ \\
\hline
AIC                & 675.59                  & 665.56                  & 511.42                  \\
\hline
\hline
repetition.order.1   & $8.22 \; (0.77)^{***}$  & $8.40 \; (1.52)^{***}$  & $7.21 \; (1.49)^{***}$  \\
repetition.order.2   &                         & $-0.09 \; (0.72)$       & $-1.35 \; (0.62)^{*}$   \\
repetition.order.3   &                         &                         & $1.91 \; (0.53)^{***}$  \\
meeting.size         & $-1.56 \; (0.17)^{***}$ & $-1.56 \; (0.18)^{***}$ & $-1.68 \; (0.19)^{***}$ \\
meeting.size.squared & $1.67 \; (0.15)^{***}$  & $1.67 \; (0.15)^{***}$  & $1.72 \; (0.16)^{***}$  \\
\hline
AIC                  & 214.79                  & 216.78                  & 209.21                  \\
\hline
\hline
repetition.order.1   & $5.99 \; (1.42)^{***}$  & $9.47 \; (2.88)^{***}$  & $9.81 \; (3.04)^{**}$   \\
repetition.order.2   &                         & $-2.50 \; (2.00)$       & $-2.29 \; (2.42)$       \\
repetition.order.3   &                         &                         & $-0.49 \; (1.34)$       \\
repetition           & $4.38 \; (0.87)^{***}$  & $4.54 \; (0.95)^{***}$  & $4.59 \; (0.97)^{***}$  \\
meeting.size         & $-2.18 \; (0.34)^{***}$ & $-2.24 \; (0.35)^{***}$ & $-2.23 \; (0.35)^{***}$ \\
meeting.size.squared & $1.11 \; (0.16)^{***}$  & $1.14 \; (0.16)^{***}$  & $1.13 \; (0.16)^{***}$  \\
\hline
AIC                  & 99.76                   & 100.28                  & 102.11                  \\
\hline
\multicolumn{4}{l}{\scriptsize{$^{***}p<0.001$, $^{**}p<0.01$, $^*p<0.05$}}
\end{tabular}
\caption{PM meetings data analyzed by RHEM with unconstrained event size. All models have been estimated on 886 events and 9,727 observations. (Note that the number of observations is the number of events plus the number of sampled non-event hyperedges.)}
\label{table:PM_uncond_size}
\end{center}
\end{table}

\paragraph{Unconstrained risk set.} Table~\ref{table:PM_uncond_size} reports estimated models considering the unconstrained risk set. Discussing first the model fit indicator (AIC), we find that including the size and squared size of hyperevents brings the biggest improvement in model fit, followed by the improvements implied by the repetition statistic. The model fit assesses how well the model can distinguish between events and non-events from the set of observations, in the sense that ``better'' models assign higher rates to event hyperedges and lower rates to non-event hyperedges. Looking at Fig.~\ref{fig:pm_sizes} it seems very plausible that the size and squared size of hyperedges serve this purpose very well: an observation with a size close to the minimum or close to the maximum is more likely to be an event, while an observation with a size close to the median is more likely to be a non-event. This makes it plausible why including the two statistics dependent on size improve the AIC to such a large extent. The parameter associated with the square of the hyperedge size is positive (pointing to a U-shaped quadratic function that takes larger values at the extremes) and the parameter associated with the size of the hyperedge is negative -- shifting the minimum of this polynomial to the positive numbers. 

Including the repetition statistic brings the second-largest improvement to the model fit. The parameter associated with repetition is consistently positive, so that the event rate on a hyperedge $h$ is typically higher the more previous events happened on the identical hyperedge $h$. In particular, we consistently find higher-order dependence -- going beyond purely dyadic event rates -- among hyperevents. 

The effect of sub-repetition of order one is consistently positive, implying that actors that have been more active in the past (i.\,e., that have participated in more events) are more likely to be included among the participants of the next event. From another point of view, a hyperedge has a higher event rate if it is composed of participants that have been more active before.

Adding sub-repetition of order two or three on top of the other effects has an inconsistent effect on model fit and yields parameters that strongly depend on whether or not repetition or the statistics dependent on event size are included. Considering the block of the first three models we find that adding sub-repetition of order two and three to models not controlling for repetition or event size brings a modest improvement in the model fit. Parameters associated with sub-repetition of order two and three are negative in models neither controlling for hyperedge size nor for repetition. We claim that these negative parameters are spurious effects that result from the failure to control for hyperedge size. Indeed, we find that 483 events (that is more than half of all events) have a hyperedge size equal to one. If a hyperedge has a non-zero value in sub-repetition of order two or three, it is a signal that this hyperedge cannot have size one. Thus it cannot be among those observations that have the the highest event rates -- explaining why sub-repetition of order two or three has a seemingly negative effect on event rates in models not controling for event size.

We conclude that if sizes of observed events and sizes of sampled non-event hyperedges have different distributions -- as those shown in Fig.~\ref{fig:pm_sizes} -- relational hyperevent models necessarily have to control for hyperedge size. Failure to do so may lead to spurious results due to correlations of explanatory variables with (functions of) hyperedge size.

We further find that, once we control for repetition, size, and square size of hyperedges, sub-repetition of order two and three have no significant effect on event rates and including them actually decreases model fit. The model with the highest model fit is the one with AIC=99.76, including sub-repetition of order one (i.\,e., past activity of individual participants), repetition (i.\,e., past activity of the exact same hyperedge), size of hyperedges, and squared size of hyperedges. The event rate on a given hyperedge $h$ is higher if the size of $h$ is either close to the minimum or close to the maximum, if the actors in $h$ participated individually in many previous events, and/or if the exact same hyperedge experienced many previous events. On top of these effects, the number of previous shared events of dyads or triads contained in $h$ has no further effect on the event rate.

We may wonder whether the quadratic function of hyperedge size captures the effect of size on event rates sufficiently well. We experimented with estimating fixed effects for all possible meeting sizes. This, however, resulted in a non-convergent model due to the fact that some possible meeting sizes are not realized by any event in our empirical data. We leave it to future work to develop more sophisticated models for the effect of size on event rates.

\paragraph{Conditional-size models.} One way to circumvent the question how to best control for the effect of hyperedge size is to consider the observed sizes as given (rather than random) and model event rates conditional on these event sizes, as it has been discussed in Sect.~\ref{sec:conditional_size}. That is, conditional size models define the risk set associated with an observed hyperevent $e=(h_e,t_e)$ to be the set of all subsets of size $|h_e|$ of the set of actors $V_{t_e}$. Table~\ref{table:PM_cond_size} reports findings on these kind of models (in which we do not have to -- and actually cannot -- control for the effect of hyperedge size).

\begin{table}
\begin{center}
\begin{tabular}{l c c c }
\hline
repetition.order.1 & $1.24 \; (0.06)^{***}$ & $0.92 \; (0.06)^{***}$ & $0.93 \; (0.06)^{***}$ \\
repetition.order.2 &                        & $2.14 \; (0.16)^{***}$ & $0.98 \; (0.16)^{***}$ \\
repetition.order.3 &                        &                        & $2.17 \; (0.27)^{***}$ \\
\hline
AIC                & 3350.83                & 3083.56                & 3000.43                \\
\hline
\hline
repetition.order.1 & $0.85 \; (0.09)^{***}$ & $0.19 \; (0.11)\phantom{^{***}}$       & $0.37 \; (0.11)^{***}$ \\
repetition.order.2 &                        & $2.24 \; (0.16)^{***}$ & $1.32 \; (0.18)^{***}$ \\
repetition.order.3 &                        &                        & $1.82 \; (0.28)^{***}$ \\
repetition         & $0.29 \; (0.05)^{***}$ & $0.50 \; (0.06)^{***}$ & $0.38 \; (0.07)^{***}$ \\
\hline
AIC                & 3324.65                & 3019.10                & 2967.39                \\
\hline
\multicolumn{4}{l}{\scriptsize{$^{***}p<0.001$, $^{**}p<0.01$, $^*p<0.05$}}
\end{tabular}
\caption{PM meetings data analyzed by conditional-size RHEM. All models have been estimated on 886 events and 8,961 observations.}
\label{table:PM_cond_size}
\end{center}
\end{table}

In conditional-size models we find that the effects of all statistics are very consistent across models and every statistic improves the model fit when added to any baseline model. The best-fitting model, thus, is the model including repetition and sub-repetition of order one, two, and three. This model suggests that the event rate on a given hyperedge $h$ -- compared to other hyperedges of the same size as $h$ -- is typically higher if (1) the members of $h$ individually participated in many previous events, (2) each pair composed of two members of $h$ co-participated dyadwise in many common previous events, (3) each triad composed of three members of $h$ co-participated triadwise in many common previous events, and (4) all members of $h$ co-participated in many common previous events, without any further participants. In particular, we find higher-order dependence in hyperevents also with conditional-size models.

How can we interpret the partially different findings resulting from unconstrained models and conditional-size models? One way to find an interpretation is to assume hypothetical modifications of an hyperedge $h$ -- transforming it into a hyperedge $h'$ -- and discuss by which factor the event rate on $h'$ is different from the event rate on $h$, according to the fitted model. For the unconstrained models we may compare any possible hyperedge $h'$ with $h$. For instance, $h'$ may result from adding any number of actors to $h$, from removing any number of actors to $h$, or from combinations of adding and removing actors. We may for instance, consider how the event rate changes when we add a single new participant, or a set of new participants, to $h$. In conditional-size models only transformations that leave the number of participants unchanged are allowed. Thus, we may for instance consider how the event rate changes when we add a new participant $v\not\in h$ to the hyperedge $h$, at the expense of another actor $v'\in h$ that is removed at the same time as $v$ is added. Similarly we may consider how the event rate changes when we add a set of $k$ new actors to the hyperedge $h$ at the expense of removing another set of $k$ actors from $h$. Thus, conditional-size models take the point of view that individual actors -- or sets of actors of the same size -- \emph{compete} for participation in events. Specifically, assume that an actor $v\not\in h$ has more previous common events with other members of the hyperedge $h$ than another actor $v'\in h$. The conditional-size models reported in Table~\ref{table:PM_cond_size} suggest that the hyperedge $h\cup\{v\}\setminus\{v'\}$ (containing $v$ instead of $v'$) has a higher event rate than the hyperedge $h$ (containing $v'$ instead of $v$). That is, actor $v$ has a higher probability to experience a common event with the other participants of $h$ than actor $v'$. If the other participants of $h$ were the ones deciding about the composition of the next event, they would be inclined to include $v$ at the expense of $v'$. The interpretation of the revealed effects in unconstrained models is more ambiguous since not only substitution of (sets of) actors by others are allowed but also modifications that increase or decrease hyperevent sizes. Thus, the validity of effects revealed by unconstrained models hinges on the assumption that the effect of hyperedge size on event rates is perfectly captured by the model. Failure to control for size effects can lead to spurious findings.

We observe that model fit indicators of models not constraining hyperedge size are much lower than those of the conditional-size models. We emphasize, however, that it would be invalid to conclude that unconstrained models are ``better'' models than conditional-size models in any sense. Model fit indicators such as the AIC can compare different models estimated on the \emph{same} data -- but it is invalid to compare the AIC of two models that have been estimated on \emph{different} data. (Note that conditional-size models are indeed estimated on different data since the non-event hyperedges associated with every observed hyperevent are drawn from a restricted set.) The reason why unconstrained models achieve a lower AIC is that the task to identify events (or non-events) out of all given observations is much simpler when we do not condition on the size of hyperedges. As discussed above, in unconstrained models we can apply the very simple but effective criterion that observations whose size is close to the minimum or close to the maximum are more likely to be events while observations with intermediate hyperedge sizes are more likely to be non-events. From another point of view, the non-events in unconstrained models have a totally implausible distribution of event sizes, as is apparent from Fig.~\ref{fig:pm_sizes}, and can therefore be assigned low event rates, leading to good model fit. This simple criterion no longer works for conditional-size models, where events and sampled non-events have the identical distribution of hyperedge sizes by design. From another point of view, the non-events resulting from conditional-size models have a very plausible distribution of hyperedge sizes and are therefore harder to recognize as non-events.

% ----------------------------------------------------------------------
\section{Case study II: Team assembly and team performance in co-author networks}
\label{sec:coauthor}

The second empirical case study that we consider analyzes a co-author network. Hyperevents correspond to published papers, where the underlying hyperedge is given by the set of authors and the event time is the year of publication. In this data we have a relational outcome variable associated with events giving the (normalized) number of citations received by the respective paper. We analyze processes explaining team assembly (i.\,e., which authors tend to publish papers together) and team performance (i.\,e., what is the impact of papers published by different teams). Previous work has analyzed team assembly \citep{guimera2005team} and team performance \citep{ahmadpoor2019decoding} in publication networks with different methods.

% ----------------------------------------------------------------------
\subsection{Setting and data}
\label{sec:coauthor_data}

\begin{figure}
  \begin{center}
    \includegraphics[width=0.7\textwidth]{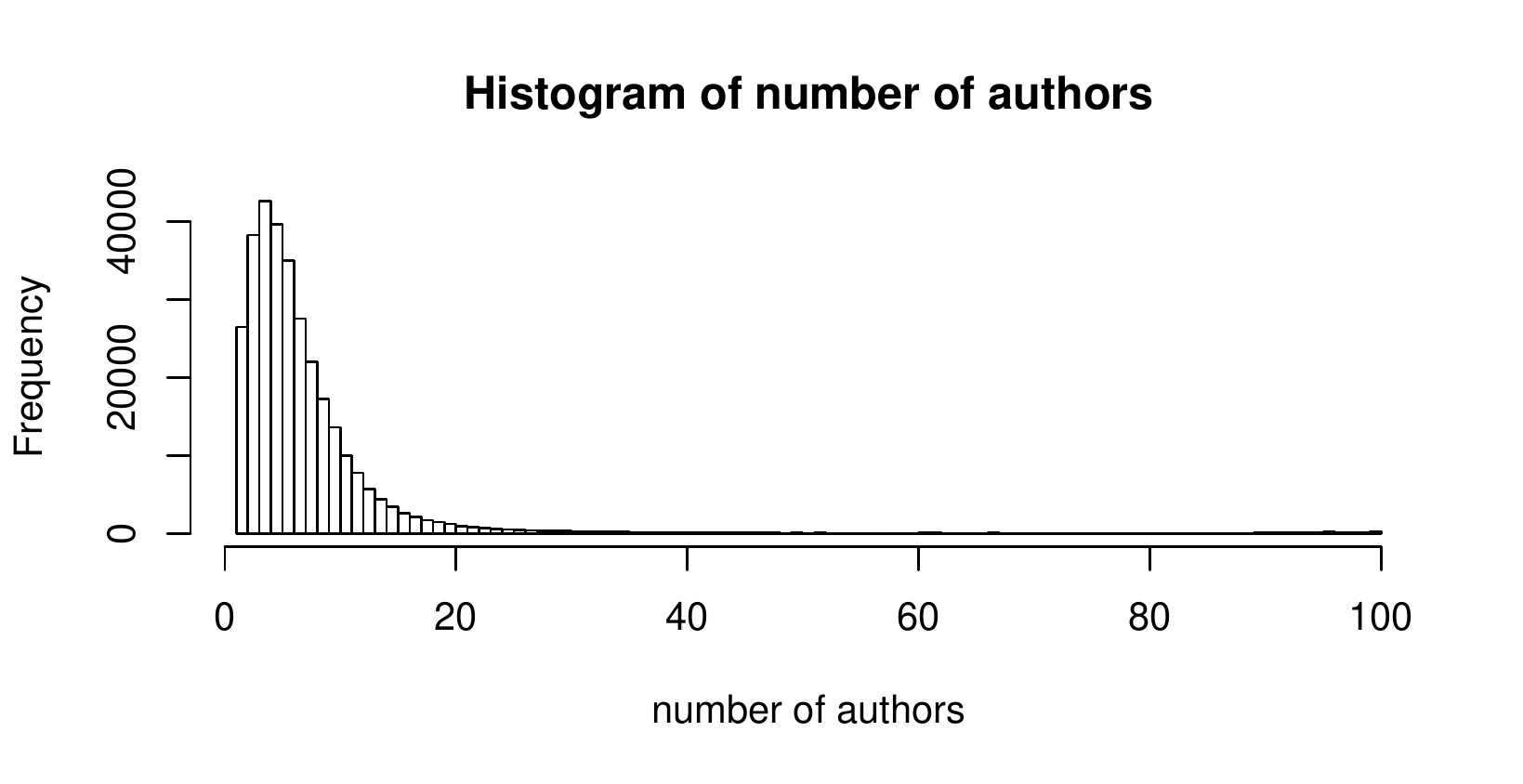}
  \end{center}
  \caption{Histogram of sizes of observed hyperevents in the co-author data. Hyperevent size is the number of authors of published papers.}
  \label{fig:coauthor_sizes}
\end{figure}

We have data on 315\,238 scientific papers among 523\,152 authors published from 1965 to 2019. The papers have been chosen by selecting from the set of EU-based researchers publishing in the social sciences the 1,200 that are most productive in terms of the number of publications (``seed authors''). Then we included all papers of these 1,200 seed authors and all papers published by any of their co-authors -- including those papers published by co-authors without any of the seed authors. We restrict the number of authors of any paper to 100 (if a paper exceeds this limit, we consider only the first 100 authors). With this cut-off, the average number of authors per paper is 7.95 and the average number of papers per author is 4.79. The histogram of the number of authors per paper (that is, of the hyperevent size) is given in Fig.~\ref{fig:coauthor_sizes} which suggests that small events have a distribution of hyperedge sizes resembling a binomial distribution but also that large events are over-represented and few events have sizes far exceeding the average. Thus, the distribution of event sizes seems to be a mixture of binomial with a heavy-tailed distribution.

% ----------------------------------------------------------------------
% ----------------------------------------------------------------------

Hyperevents in this data have an associated relational outcome which is the number of citations that the paper has received by the time of data collection in 2019. The number of citations is heavily right-skewed with a median of eight, a mean equal to 25.7, maximum of 10\,687, and standard deviation equal to 87.7. We normalize this count by subtracting the average number of citations of papers published in the same year. The resulting \emph{performance}, thus, gives the excess number of citations. It is a positive number if the paper attracted more citations than the average paper published in the same year and it is negative if it attracted less citations than expected, given the year of publication. Citing papers may be published at any time until the time of data collection. In this case study, we consider the excess number of citations as an indication of performance of the team of authors and assume that this performance (that is, the quality or excellence of the paper) is transparent to other actors in the year following the publication of the paper so that it might, for instance, influence team selection of papers published afterwards. We leave a more refined model which considers also the timing of citations to future work.

Publication data, as the example we consider in this case study, usually comes with a relative coarse time granularity given by the year. Thus, necessarily many events happen \emph{simultaneously}, that is, they have the same time stamp. In this case study we assume that simultaneous events are conditionally independent of each other, given the history of previous events. Thus, event rates (or relational outcome variables) on given hyperedges may depend on \emph{previous} events on the same or other hyperedges, but are assumed to be conditionally independent of the occurrence or non-occurrence of simultaneous events. Previous work has shown that the validity of the assumption of conditional independence in periodically observed dynamic networks depends on the spacing of observation times \citep{lerner2013conditional}. We leave it to future work to develop a model that can cope with mutual dependence among simultaneous events.

% ----------------------------------------------------------------------
\subsection{Model specification}
\label{sec:coauthor_models}

In this case study we are interested in effects explaining team assembly and team performance. More specifically, we first want to analyze how team assembly depends on familiarity and prior performance on various levels. Among the familiarity effects we test whether actors have the tendency to repeat collaboration dyadwise, triadwise, or within the identical team. Dependence on prior performance would induce actors to seek collaboration with others who individually performed well in the past, or with whom they have a history of prior \emph{shared} success -- dyadwise, triadwise, or within the identical team. Secondly, within the family of relational outcome models, we seek to explain team performance (that is, the excess number of citations received by papers) by familiarity and prior performance on various levels. In particular, we are interested in whether effects explaining team assembly and team performance are ``consistent'' in the sense that actors have a preference to collaborate within teams that are likely to produce high-impact work. Team assembly is modeled with Cox-proportional hazard models given in Eq.~(\ref{eq:sampled_likelihood}) and team performance is modeled with relational outcome models given in Eq.~(\ref{eq:likelihood_rom}) where the distribution $f$ of team performance is chosen to be the normal distribution, leading to linear regression (that is, ordinary least squares).

Statistics operationalizing assumed effects of familiarity on team assembly and team performance are identical with the effects considered in the previous case study on meetings: repetition, sub-repetition of order one, two, and three, as well as the hyperedge size (though in the co-author model we do not use the square of the hyperedge size). Statistics operationalizing assumed effects of prior (shared) performance on team assembly and team performance are similar but take relational outcome (that is, excess number of citations) of previous events into account. Specifically, we define two attributes for hyperedges $h$. The first measures aggregated performance resulting from previous events occuring on the identical hyperedge:
\[
\textit{hyperedge.performance}(h;t;E)=\sum_{e\in E_{<t}}y_e\cdot\chi(h = h_e)\enspace,
\]
where $y_e$ is the relational outcome of the hyperevent $e$ and $\chi(\cdot)$ is the indicator function that is one if the argument is true and zero else. 
The second attribute measures performance resulting from previous events occuring on any hyperedge containing $h$: 
\[
\textit{sub-hyperedge.performance}(h;t;E)=\sum_{e\in E_{<t}}y_e\cdot\chi(h\subseteq h_e)\enspace.
\]
Based on these attributes, we define the statistic
\[
\textit{prior.hyperedge.success}(h;t;G[E;t])=\frac{\textit{hyperedge.performance}(h;t;E)}{\textit{hyperedge.activity}(h;t;E)}\enspace,
\]
to assess the average prior success of events occuring on the identical hyperedge and for $p=1,2,3$ we define
\[
\textit{prior.sub-hyperedge.success}^{(p)}(h;t;G[E;t])=
\frac{\sum_{h'\in\sub^{(p)}(h)}\textit{sub-hyperedge.performance}(h';t;E)}
     {\sum_{h'\in\sub^{(p)}(h)}\textit{hyperedge.degree}(h';t;E)}\enspace,
\]
to assess the average prior success of events in which all members of any sub-hyperedge of order $p$ (and potentially further actors) jointly participated.\footnote{We resolve $\frac{0}{0}=0$ in both definitions.} 

In this much larger network, we do not consider unconstrained models that sample from the full space of hyperedges of all possible sizes. Indeed, such models would produce samples of non-events that have hyperedge sizes by many orders of magnitude larger than that of observed events. (A random hyperedge drawn uniformly from all subsets of the set of authors would have on average more than 250\,000 participants. Thus, the full risk set would have an absurd distribution of hyperedge sizes which is totally different from the distribution of observed event sizes.) We explore two model variants whose risk sets have a plausible distribution of event sizes: (1) conditional-size models, which have been already applied in the previous case study in Sect.~\ref{sec:meetings} and (2) models analyzing only \emph{repeated} events and therefore restrict the risk set at time $t$ to those hyperedges that have experienced at least one event before $t$. The second type of models, restricting the analysis to repeated events, is complemented by conditional-size models restricted to \emph{first} events, that is, restricted to events occurring on hyperedges that have never experienced an event before.

The need to analyze first events and repeated events separately, becomes apparent from a preliminary, exploratory analysis. If we try to fit conditional-size models that include the repetition statistic to all events (first events and repeated events) we typically make the experience that model estimation does not converge and that during the course of the estimation the repetition parameter diverges towards infinity. This observation becomes understandable when we look at some characteristics of the co-author data. We find that about 11.6\% of all events are repeated events, that is they occur on hyperedges $h$ with $\textit{repetition}(h;t;E)>0$. From another point of view, even though the repeated events are not distributed uniformly, a hyperedge that has experienced at least one event has a non-negligible probability to experience another event in the future. On the other hand, a randomly chosen hyperedge of moderately large size that has never experienced an event before has a probability numerically indistinguishable from zero to ever experience any event in the future. For instance, the total number of hyperevents (which is about 300\,000) is numerically as good as zero compared to the more than $500\,000 \choose 100$ hyperedges of size 100. Thus, the simple binary indicator $\chi(\textit{repetition}(h;t;E)>0)$ increases the event rate on hyperedges by a factor that is numerically as good as infinity -- which is why model estimation either does not converge or, if it converges, is very unreliable and typically has a high variability over different samples. This problem seems to be related with -- but is much more severe than -- the unreliability of estimating the repetition parameter in large \emph{dyadic} event networks reported in \citet{lerner2019reliability}. We recall that the risk set size in the hyperevent network considered in this section is by orders of magnitude larger than the risk set size considered in \citet{lerner2019reliability}.

We emphasize that the issue sketched in the preceding paragraph is not just a technical problem of model estimation, but rather points to a conceptual problem. Even if we were given unlimited computational power, could compute the full likelihood function without case-control sampling given in Eq.~(\ref{eq:likelihood}), and could represent numbers with arbitrarily many digits, the rate-increase implied by $\chi(\textit{repetition}(h;t;E)>0)$ would still be practically as good as infinity and could potentially mask or distort other -- much weaker, but still significant and relevant -- effects in hyperevent networks. Based on our current limited experience with RHEM, it seems that we have to accept that first events and repeated events are generated by different processes and therefore have to be analyzed by separate models in networks whose size exceeds a moderate number of nodes.

% ----------------------------------------------------------------------
\subsection{Results}
\label{sec:coauthor_results}

\paragraph{Team assembly.} Table~\ref{table:coauthor_rate} reports estimated parameters of models explaining team assembly, that is, the relative event rate on hyperedges. As discussed above, we estimate two separate models, one explaining first events on any hyperedge and the second explaining repeated events. The first model does not include the statistics repetition and past performance of the exact hyperedge (since these statistics are constantly zero on all instances considered in this model) and does not include the number of authors (since the model conditions on the size of observed hyperevents). To construct the sampled likelihood given in Eq.~(\ref{eq:sampled_likelihood}), we sample one control (non-event hyperedge) for every observed event. Following recommendations from \citet{lerner2019reliability} we repeat sampling ten times and fit models separately to the different samples. Findings are very reliable across samples in the sense that all parameters that are significantly different from zero in Table~\ref{table:coauthor_rate} have the identical sign across all ten samples. 

\begin{table}
\begin{center}
\begin{tabular}{l r r }
\hline
 & First events (conditional size) & Repeated events\\
\hline
repetition.order.1         & $0.371 \; (0.001)^{***}$  & $-0.060 \; (0.007)^{***}$ \\
repetition.order.2       & $0.058 \; (0.001)^{***}$  & $0.039 \; (0.006)^{***}$  \\
repetition.order.3         & $-0.004 \; (0.001)^{**\phantom{*}}$  & $0.087 \; (0.005)^{***}$  \\
repetition                     &                           & $0.099 \; (0.004)^{***}$  \\
prior.sub-hyperedge.success.order.1      & $-0.223 \; (0.004)^{***}$ & $-0.132 \; (0.009)^{***}$ \\
prior.sub-hyperedge.success.order.2      & $0.028 \; (0.001)^{***}$  & $0.035 \; (0.012)^{**\phantom{*}}$   \\
prior.sub-hyperedge.success.order.3      & $0.040 \; (0.001)^{***}$  & $0.016 \; (0.008)\phantom{^{***}}$        \\
prior.hyperedge.success &                           & $0.006 \; (0.009)\phantom{^{***}}$        \\
number.of.authors            &                           & $-2.169 \; (0.027)^{***}$ \\
\hline
AIC                            & 5558056.916               & 570139.996                \\
Num. events                    & 278,527                    & 36,708                     \\
Num. obs.                      & 566,601                    & 63,860                     \\
\hline
\multicolumn{3}{l}{\scriptsize{$^{***}p<0.001$, $^{**}p<0.01$, $^*p<0.05$}}
\end{tabular}
\caption{Estimated parameters of RHEM modeling the relative rate of co-author events (that is, publication rates) on hyperedges. \textit{Left}: analysis restricted to first events, that is, events whose exact hyperedge has never experienced an event before. Risk set associated with event $e$ are all hyperedges of the same size as $h_e$ that have never experienced an event before. \textit{Right}: analysis restricted to repeated events, that is, events whose exact hyperedge has experienced at least one previous event. The risk set associated with event $e$ contains all hyperedges that have experienced an event before $t_e$.}
\label{table:coauthor_rate}
\end{center}
\end{table}

We find that in most cases actors have a tendency to repeat established collaboration, that is, to choose familiar actors as team members for future publications. In particular, the rate of first events (that is, publications whose authors have never published a paper with the identical set of authors before) on a hyperedge $h$ typically gets higher with the average number of previous publications of all individual authors in $h$ and the average number of dyadwise co-authored papers of all pairs of authors in $h$ but it typically decreases with the average number of triadwise co-authored papers of all triads of authors in $h$ (although the latter effect is rather small). When analyzing repeated events (that is, publications  whose authors have published at least one paper with the identical set of authors before), we find that the rate of repeated events on a hyperedge $h$ typically gets higher with growing numbers of dyadwise and triadwise co-authored papers and also gets higher with a growing number of previous papers involving the identical set of authors $h$. In contrast, we find that the average number of previous publications of all individual authors in $h$ tends to decrease the rate of repeated events. Thus, actors seem to have a preference for repeating collaboration with familiar others, but -- controlling for these patterns -- there seems to be a saturation effect on individual publication activity. When analyzing repeated events, we find that larger hyperedges (that is, hyperedges with more authors) have smaller publication rates -- controlling for all other effects.

Regarding the effect of prior success (received citations of previously published papers) on publication rates, we find that the publication rate on a hyperedge $h$ decreases with growing average prior success of the individual members of $h$ in both models (that is, for first events and repeated events). Thus, previously successful authors seem to publish less in the future -- controlling for all other effects. In contrast to this pattern, we find that the rate of first events on hyperedges $h$ tends to increase with prior dyadwise and triadwise \emph{shared} success, while the rate of repeated events is only increased by prior dyadwise shared success but not significantly influenced by prior triadwises shared success. The effect of prior shared success of the hyperedge $h$ is also not significant for the publication rate on the identical hyperedge $h$.

\paragraph{Team performance.} How do the same characteristics that influence team assembly (publication rates on hyperedges) impact the performance of future publications? Table~\ref{table:coauthor_perf} reports estimated parameters of relational outcome models that explain the performance (excess number of citations of papers, given the year of publication) by the same statistics used in models for team assembly.

\begin{table}
\begin{center}
\begin{tabular}{l r r }
\hline
 & First events  & Repeated events\\
\hline
(Intercept)                    & $1.059 \; (0.163)^{***}$  & $-8.033 \; (0.311)^{***}$  \\
repetition.order.1         & $-0.389 \; (0.205)\phantom{^{***}}$       & $1.646 \; (0.375)^{***}$   \\
repetition.order.2       & $1.151 \; (0.321)^{***}$  & $-1.116 \; (0.441)^{*\phantom{**}}$    \\
repetition.order.3         & $-4.757 \; (0.299)^{***}$ & $1.520 \; (0.363)^{***}$   \\
repetition                     &                           & $0.529 \; (0.355)\phantom{^{***}}$         \\
prior.sub-hyperedge.success.order.1      & $7.214 \; (0.185)^{***}$  & $0.130 \; (0.436)\phantom{^{***}}$         \\
prior.sub-hyperedge.success.order.2      & $4.940 \; (0.236)^{***}$  & $28.176 \; (0.630)^{***}$  \\
prior.sub-hyperedge.success.order.3      & $9.013 \; (0.221)^{***}$  & $-12.166 \; (0.430)^{***}$ \\
prior.hyperedge.success &                           & $17.901 \; (0.468)^{***}$  \\
number.of.authors            & $8.276 \; (0.184)^{***}$  & $-0.923 \; (0.326)^{**\phantom{*}}$   \\
\hline
R$^2$                          & 0.051                     & 0.274                      \\
Adj. R$^2$                     & 0.051                     & 0.274                      \\
Num. obs.                      & 278,527                    & 36,708                      \\
RMSE                           & 86.159                    & 59.663                     \\
\hline
\multicolumn{3}{l}{\scriptsize{$^{***}p<0.001$, $^{**}p<0.01$, $^*p<0.05$}}
\end{tabular}
\caption{Modeling the relational outcome (``success'') of co-authored papers, that is, the difference between the number of citations of the paper and the expected number of citations, given the year of publication. \textit{Left}: analysis restricted to first events, that is, events whose exact hyperedge has never experienced an event before. \textit{Right}: analysis restricted to repeated events, that is, events whose exact hyperedge has experienced at least one previous event.}
\label{table:coauthor_perf}
\end{center}
\end{table}

We find (with few exceptions) a rather consistent pattern that the number of citations of a published paper typically gets larger with growing prior success of individual authors as well as with growing prior shared success of dyads, triads, and the identical team of authors. An exception is the negative effect of triadwise prior shared success on the success of repeated events; prior performance of individual authors is non-significant for the success of repeated events. In contrast to prior (shared) success, familiarity of authors seems to have a less consistent impact on publication success.
The previous number of publications of individual members of the hyperedge tends to increase the number of received citations of repeated events but is insignificant for the success of first events. 
The number of previously dyadwise co-authored papers has a positive effect on the expected success of first events but a negative effect on the success of repeated events. This pattern is reversed for the effect of triadic familiarity. 
In the model for repeated events we find that the number of previous publications of the identical team of authors has no significant impact on team performance. The number of authors of a paper (that is, the team size) positively impacts the success of first events but is slightly negative for the success of repeated events.

In summary, team assembly seems to be mostly explained by familiarity so that the event rate on a hyperedge $h$ tends to be higher if the participants of $h$ have collaborated before. In contrast, team performance seems to be mostly explained by prior shared success of participants. Models discussed in this section also reveal higher-order dependence in empirical co-author networks which cannot be modeled by purely dyadic publication rates.

% ----------------------------------------------------------------------
\section{Discussion}
\label{sec:discussion}

In this section we first discuss general methodological insight derived from the two empirical case studies and then provide a detailed discussion of previous work that is most related to our model in Sect.~\ref{sec:related}.

From a high-level view, RHEM that model the rate of hyperevents compare hyperedges experiencing events at given points in time (``cases'') with hyperedges from the risk set that could potentially have experienced an event at that time but did not (``controls,'' ``alternatives,'' or ``non-events''); see for instance Eq.~(\ref{eq:sampled_likelihood}). Models seek to distinguish the event hyperedges from the alternative hyperedges by assigning high event rates to the former and low event rates to the latter. A recurrent insight from the empirical case studies -- which also seems to point to a major challenge in further developing and understanding RHEM -- is the importance of defining \emph{plausible} alternative hyperedges for observed events. One of the most crucial aspects influencing the plausibility that hyperedges might experience events -- although not the only one -- is their size.

Our first case study revealed that models that neither constrain the size of alternative hyperedges nor control for the effect of size on event rates (such as the first three models reported in Table~\ref{table:PM_uncond_size}) are completely useless. The distribution of hyperedge size in the unconstrained risk set is extremely different from that of the observed events (compare Fig.~\ref{fig:pm_sizes}). Failure to control for the effect of size on the event rate can lead to spurious effects for any statistic correlating with (functions of) size. In the first case study we considered two model variants to deal with the effect of hyperedge size: (1) including size and the squared size among the model statistics (which was motivated by the observed U-shaped distribution of the size of observed events) and (2) conditioning on hyperedge size, that is, comparing each hyperedge experiencing an event only with alternatives of the same size. The empirical case study from Sect.~\ref{sec:meetings} could not answer conclusively which of the two possibilities is preferable in general. However, in the given study, conditional-size models produced more consistent results and also turned out to be interpretable in a more straightforward way. In general, analysts face a trade-off between constraining the risk set not enough and constraining it too much. An unconstrained risk set could yield implausible alternative hyperedges while a risk set that is too heavily constrained might lead to models conditioning on some aspects of hyperevents that in reality result from an endogenous process (for example, hyperedge size results from the choice to include actors in hyperevents).

The second case study revealed that conditioning on hyperedge size might still be insufficient in larger hyperevent networks. Indeed, given many actors nearly every randomly drawn hyperedge is an ``implausible'' alternative candidate for the next hyperevent (see the discussion in Sect.~\ref{sec:coauthor_models}). In such models we can often find very simple criteria (such as the property of having previously experienced at least one event) that increase the event rate on hyperedges by factors which are numerically indistinguishable from infinity. Based on our current limited experience with RHEM, we recommend to partition the whole population of hyperedges along such criteria that have a dramatic (and typically obvious) effect on event rates. In the concrete case study from Sect.~\ref{sec:coauthor}, we specify separate models for \emph{first} events (considering hyperedges that have never experienced any event before) and for \emph{repeated} events (considering hyperedges that have experienced at least one event before). In general, other criteria might additionally be used to split the population of hyperedges into more homogeneous subsets. Such criteria might, for instance, distinguish hyperedges in which every pair of actors have experienced at least one common event, hyperedges in which every pair of actors have co-participated in an event with at least one common other third actor, or hyperedges that are homogeneous with respect to an exogenously given covariate, e.\,g., institutional affiliation.

% ----------------------------------------------------------------------
\subsection{Related work}
\label{sec:related}

In this section we discuss some of the most-related previous work in detail. In particular, we emphasize differences between models proposed in previous work and RHEM.

\citet{guimera2005team} propose a model in which teams consisting of any number of actors -- representing for instance, the set of co-authors of a scientific paper -- are randomly assembled by including team members one by one. The probability of an actor $A$ to be included in team $T$ can depend (1) on characteristics of the actor $A$, such as being an incumbent or a newcomer and (2) on the existence of previous collaboration of $A$ with actors who have previously been selected into the growing team $T$. Thus, this model can incorporate dyadic dependence, but no higher-order dependence. For instance, assume a situation in which actor $A$ has frequently collaborated with actors $B$ and $C$, but never with both together -- perhaps because $A$'s papers co-authored with $B$ are from a different area than $A$'s papers co-authored with $C$. In this situation we could expect an increased probability that teams contain the dyad $\{A,B\}$ and an increased probability that teams contain the dyad $\{A,C\}$ -- but at the same time a decreased probability that a team contains the triad $\{A,B,C\}$. Such a higher-order dependence, however, cannot be specified in the model from \citet{guimera2005team}. Indeed, in their model $B$ and $C$ would both be more likely to be in a team with $A$ and thus -- by chance alone -- the triad $\{A,B,C\}$ would be over-represented.

\citet{b-refsa-08} discussed the possibility to represent several senders or receivers by new dummy nodes representing sets of actors: '\emph{To treat simultaneous joint action by multiple senders and/or receivers, we simply create one or more ``virtual'' senders and/or receivers that represent subsets of the original sender/receiver set.}' \citep[][p.159]{b-refsa-08}. Examples of such virtual receivers include ``broadcast'' actors representing, for instance, a whole school class which may receive broadcast messages from the teacher \citep{dubois2013hierarchical}. It is easy to see that the strategy to represent sets of senders or receivers by dummy nodes is only possible for a limited number of predefined sets of actors. Creating these dummy nodes for all subsets would result in a prohibitively huge graph with an exponential number of nodes in which most nodes never participate in any event. Moreover, the inclusion relation among these virtual nodes representing subsets would have to be incorporated somehow in the model. Indeed if some set of actors $S$ receives a message then all subsets $S'\subset S$ are also receiving that message -- which may influence the future rate of events to or from $S'$.

\citet{kim2018hyperedge} propose the hyperedge event model (HEM) for directed events that have exactly one sender but can have any number of receivers (or conversely any number of senders and exactly one receiver). Their model specifies for every possible sender-receiver pair $(i,j)$ an intensity function $\lambda_{ij}$. These dyadic intensity functions then stochastically determine (1) who is the sender of the next event and (2) which is the set of receivers, given the chosen sender. The difference to the RHEM proposed in our paper is that in the model from \citet{kim2018hyperedge} the intensities are defined for dyads, while RHEM can define intensities (event rates) for hyperedges of any size. Assume, for instance, that $A$ has often send emails to $B$ and to $C$ in the past, but never the same email to $B$ and $C$ -- perhaps because $B$ is a work colleague of $A$ but $C$ is a friend. Due to many past events, the dyadic intensities $\lambda_{AB}$ and $\lambda_{AC}$ would be high and, in the model from \citet{kim2018hyperedge}, by chance alone, $B$ and $C$ would have an increased probability to be in the receiver set of an email send by $A$ -- although in reality they would have a very low probability to jointly receive the same email from $A$.

\paragraph{Related work from the area of machine learning.} Machine learning considers the task of \emph{subset prediction} (or \emph{sequential subset prediction}) which seems to be strongly related to modeling hyperevents. Subset prediction seeks to predict, for instance, baskets of items that a particular customer is likely to select in the next purchase. We note that \emph{prediction} (which is a common objective in machine learning) is a different goal than \emph{explanation} \citep{breiman2001statistical,shmueli2010explain}. Data modeling often seeks to identify and/or test effects that \emph{explain} aspects of the data. (For instance, in this paper we want to assess if and how prior shared success impacts team assembly in co-authoring networks.) Machine learning, on the other hand, typically seeks to \emph{predict} data out-of-sample, such as ``which basket of items will a customer most likely buy in the future.'' The number of parameters in models applied in machine learning is typically by several orders of magnitude larger than in explanatory modeling and models applied in machine learning often use dedicated estimation techniques that prevent overfitting. The resulting models perform more like ``black boxes'': they make good predictions but do not reveal specific effects in the data generating process. RHEM proposed in this paper have been designed with the goal of explaining data (rather than predicting data) in mind. Having these differences in mind, we overview related work on subset predictions since the two areas might nevertheless learn from each other.

\citet{benson2018discrete} propose a model in which the ``utility'' of a subset is composed of (1) the sum of the utilities of the individual items of the subset plus an additional ``corrective'' utility, positive or negative, which may be specified for a given budget of $k$ subsets. Thus, this model can in principle express dependence of arbitrarily high order -- but only for a predefined set of subsets. 

\citet{benson2018sequences} propose a model to predict the next set in a sequence of sets. To predict the next set $S_{k+1}$, given the preceding sets $S_1,\dots,S_k$, their model conditions on the size of the next set and takes the union of random subsets of the preceding sets $S_i$ for $i=1,\dots,k$ until the newly constructed set is of the given size. When selecting subsets of the preceding sets $S_i,\,i=1,\dots,k$, the more recent sets typically have a higher probability to be partially repeated. More precisely, they propose to select a subset of the set $S_i$ with probability proportional to a ``recency weight'' $w_{k-i+1}$, where these recency weights are learned to maximize predictive performance. The model from \citet{benson2018sequences} can express dependencies of event rates on subset of any order and is -- from a certain point of view -- related to RHEM specified by including sub-repetition of any feasible order. Differences include that free parameters in such a RHEM are associated with subset size, while parameters in the model from \citet{benson2018sequences} are associated with lags in the sequence of sets.

% ----------------------------------------------------------------------
\section{Conclusion and future work}
\label{sec:conclusion}

In this paper, we introduce relational hyperevent models (RHEM) as a generalization of relational event models (REM) to multi-actor interaction. RHEM can specify event rates on the full space of directed or undirected hyperedges involving any number of actors. We illustratively apply RHEM in two empirical case studies on meetings of government ministers and co-authored scientific papers. These case studies provide evidence for dependencies in hyperevent networks that cannot be modeled by purely dyadic event rates but that involve hyperedges of higher order. Besides RHEM explaining event rates on hyperedges, we also define relational outcome models (ROM) explaining outcome or success resulting from given hyperevents. For instance, in a study of co-author networks, RHEM can analyze team assembly, explaining which teams of authors publish together, and ROM can analyze team performance, explaining the impact of published papers.

The computational intractability of the full likelihood function, which scales exponentially in the number of nodes, can be tackled by applying established sampling techniques. Applying case-control sampling, we succeeded in reliably estimating RHEM parameters from a network comprising hundreds of thousands of nodes and hyperevents.

From a high-level view, RHEM compare hyperedges experiencing events at given points in time (``cases'') with hyperedges from the risk set that could potentially have experienced an event at that time but did not (``controls,'' ``alternatives,'' or ``non-events''). Models seek to distinguish the event hyperedges from the alternative hyperedges by assigning high event rates to the former and low event rates to the latter. One of the major challenges in the future development of RHEM is to find appropriate definitions for \emph{plausible} alternative hyperedges for observed events. Considering the full unconstrained risk set of all hyperedges of any size leads to ``straw man'' alternatives which are completely different from any event hyperedge and can easily be recognized as non-events by the simplest of models. A minimal requirement seems to be that models control for -- or condition upon -- the size of hyperedges. In larger networks, however, conditioning on hyperedge size might still be insufficient to produce plausible alternatives. The set of all hyperedges of a given size is often extremely inhomogeneous with respect to event rates. Simple indicators, like the criterion whether a hyperedge has experienced any previous event, imply an increase in the event rate which is practically or numerically as good as infinity. Such extremely strong (and typically fairly obvious) effects could potentially mask or distort other effects of interest. Moreover, it is likely that different network effects explain events on such incomparable instances. In this paper, we therefore suggest to split the population of hyperedges along indicators that have an extreme impact on the event rate. For instance, we suggest to specify and fit separate models for new events and for repeated events. Future work is needed to better understand, and/or find alternative ways to cope with, extreme inhomogeneity over the space of hyperedges.

Another avenue for future work is to better understand hyperedge statistics and the interpretation of associated parameters. Hyperedge statistics proposed in this paper are inspired by statistics commonly applied in the specification of REM, such as repetition, reciprocation, degree effects, triadic closure, or covariate effects. However, possibilities to define hyperedge statistics are much more numerous since many statistics can be parameterized by the order of subsets involved in their definition. Some families of statistics, such as sub-repetition of various order, are nested and interrelated by design which renders the joint interpretation of associated parameters difficult. More experience in applying and interpreting RHEM in different application settings is needed.

% ----------------------------------------------------------------------
%\bibliographystyle{asa}
%\bibliography{references}

\begin{thebibliography}{23}
\newcommand{\enquote}[1]{``#1''}
\expandafter\ifx\csname natexlab\endcsname\relax\def\natexlab#1{#1}\fi

\bibitem[{Ahmadpoor and Jones(2019)}]{ahmadpoor2019decoding}
Ahmadpoor, M. and Jones, B.~F. (2019), \enquote{Decoding team and individual
  impact in science and invention,} \textit{Proceedings of the National Academy
  of Sciences}, 201812341.

\bibitem[{Benson et~al.(2018{\natexlab{a}})Benson, Kumar, and
  Tomkins}]{benson2018discrete}
Benson, A.~R., Kumar, R., and Tomkins, A. (2018{\natexlab{a}}), \enquote{A
  discrete choice model for subset selection,} in \textit{Proceedings of the
  Eleventh ACM International Conference on Web Search and Data Mining}, ACM,
  pp. 37--45.

\bibitem[{Benson et~al.(2018{\natexlab{b}})Benson, Kumar, and
  Tomkins}]{benson2018sequences}
Benson, A.~R., Kumar, R., and Tomkins, A. (2018{\natexlab{b}}), \enquote{Sequences of sets,} in \textit{Proceedings
  of the 24th ACM SIGKDD International Conference on Knowledge Discovery \&
  Data Mining}, ACM, pp. 1148--1157.

\bibitem[{Borgan et~al.(1995)Borgan, Goldstein, and
  Langholz}]{bgl-mascdcphm-95}
Borgan, {\O}., Goldstein, L., and Langholz, B. (1995), \enquote{Methods for the
  analysis of sampled cohort data in the {C}ox proportional hazards model,}
  \textit{The Annals of Statistics}, 1749--1778.

\bibitem[{Brandes et~al.(2009)Brandes, Lerner, and Snijders}]{bls-ness-09}
Brandes, U., Lerner, J., and Snijders, T.~A. (2009), \enquote{Networks Evolving
  Step by Step: Statistical Analysis of Dyadic Event Data,} in \textit{Proc.\
  2009 Intl.\ Conf.\ Advances in Social Network Analysis and Mining (ASONAM)},
  IEEE, pp. 200--205.

\bibitem[{Breiman et~al.(2001)}]{breiman2001statistical}
Breiman, L. et~al. (2001), \enquote{Statistical modeling: The two cultures,}
  \textit{Statistical science}, 16, 199--231.

\bibitem[{Bretto(2013)}]{bretto2013hypergraph}
Bretto, A. (2013), \textit{Hypergraph theory}, Cham: Springer.

\bibitem[{Butts(2008)}]{b-refsa-08}
Butts, C.~T. (2008), \enquote{A relational event framework for social action,}
  \textit{Sociological Methodology}, 38, 155--200.

\bibitem[{Chodrow(2019)}]{chodrow2019configuration}
Chodrow, P.~S. (2019), \enquote{Configuration Models of Random Hypergraphs and
  their Applications,} \textit{arXiv preprint arXiv:1902.09302}.

\bibitem[{Cox(1972)}]{cox1972regression}
Cox, D. (1972), \enquote{Regression Models and Life-Tables,} \textit{Journal of
  the Royal Statistical Society. Series B (Methodological)}, 34, 87--22.

\bibitem[{DuBois et~al.(2013)DuBois, Butts, McFarland, and
  Smyth}]{dubois2013hierarchical}
DuBois, C., Butts, C.~T., McFarland, D., and Smyth, P. (2013),
  \enquote{Hierarchical models for relational event sequences,} \textit{Journal
  of Mathematical Psychology}, 57, 297--309.

\bibitem[{Guimera et~al.(2005)Guimera, Uzzi, Spiro, and
  Amaral}]{guimera2005team}
Guimera, R., Uzzi, B., Spiro, J., and Amaral, L. A.~N. (2005), \enquote{Team
  assembly mechanisms determine collaboration network structure and team
  performance,} \textit{Science}, 308, 697--702.

\bibitem[{Kim et~al.(2018)Kim, Schein, Desmarais, and
  Wallach}]{kim2018hyperedge}
Kim, B., Schein, A., Desmarais, B.~A., and Wallach, H. (2018), \enquote{The
  Hyperedge Event Model,} \textit{arXiv preprint arXiv:1807.08225}.

\bibitem[{Lawless(2003)}]{l-smmld-03}
Lawless, J.~F. (2003), \textit{Statistical Models and Methods for Lifetime
  Data}, Wiley.

\bibitem[{Lerner et~al.(2013{\natexlab{a}})Lerner, Bussmann, Snijders, and
  Brandes}]{lbsb-mftien-13}
Lerner, J., Bussmann, M., Snijders, T.~A., and Brandes, U.
  (2013{\natexlab{a}}), \enquote{Modeling Frequency and Type of Interaction in
  Event Networks,} \textit{Corvinus Journal of Sociology and Social Policy}, 4,
  3--32.

\bibitem[{Lerner et~al.(2013{\natexlab{b}})Lerner, Indlekofer, Nick, and
  Brandes}]{lerner2013conditional}
Lerner, J., Indlekofer, N., Nick, B., and Brandes, U. (2013{\natexlab{b}}),
  \enquote{Conditional independence in dynamic networks,} \textit{Journal of
  Mathematical Psychology}, 57, 275--283.

\bibitem[{Lerner and Lomi(2019)}]{lerner2019reliability}
Lerner, J. and Lomi, A. (2019), \enquote{Reliability of relational event model
  estimates under sampling: how to fit a relational event model to 360 million
  dyadic events,} \textit{Network Science}, (to appear) available online:
  \url{https://doi.org/10.1017/nws.2019.57}.

\bibitem[{{Margaret Thatcher Foundation}(2019)}]{mtf2019}
{Margaret Thatcher Foundation} (2019), \textit{Large scale document archive,}
  retrieved from \url{https://www.margaretthatcher.org/archive}.

\bibitem[{Perry and Wolfe(2013)}]{perry2013point}
Perry, P.~O. and Wolfe, P.~J. (2013), \enquote{Point process modelling for
  directed interaction networks,} \textit{Journal of the Royal Statistical
  Society: Series B (Statistical Methodology)}, 75, 821--849.

\bibitem[{Shmueli(2010)}]{shmueli2010explain}
Shmueli, G. (2010), \enquote{To explain or to predict?} \textit{Statistical
  science}, 25, 289--310.

\bibitem[{Stadtfeld and Block(2017)}]{sb-iat-17}
Stadtfeld, C. and Block, P. (2017), \enquote{Interactions, Actors, and Time:
  Dynamic Network Actor Models for Relational Events,} \textit{Sociological
  Science}, 4, 318--352.

\bibitem[{Therneau and Grambsch(2013)}]{therneau2013modeling}
Therneau, T.~M. and Grambsch, P.~M. (2013), \textit{Modeling survival data:
  extending the {C}ox model}, Springer Science \& Business Media.

\bibitem[{Vu et~al.(2015)Vu, Pattison, and Robins}]{vpr-remslm-15}
Vu, D., Pattison, P., and Robins, G. (2015), \enquote{Relational event models
  for social learning in MOOCs,} \textit{Social Networks}, 43, 121--135.

\end{thebibliography}
% ----------------------------------------------------------------------

\end{document}